\documentclass[11pt,a4paper]{article} 
\pdfoutput=1
\usepackage{jcappub} 
\usepackage{graphicx}
\usepackage{wasysym}
\usepackage{natbib}

\title{Reconstruction of the  Dark Energy equation  of state}

\author[a,b]{J. Alberto V\'azquez}
\author[a,b]{M.~Bridges}
\author[b]{M.P.~Hobson}
\author[a,b]{A.N.~Lasenby}

\affiliation[a]{Kavli Institute for Cosmology, Madingley Road, Cambridge CB3 0HA, UK.}
\affiliation[b]{Astrophysics Group, Cavendish Laboratory, JJ Thomson Avenue, Cambridge CB3 0HE, UK.}
 
\emailAdd{jv292@cam.ac.uk}
\emailAdd{mb435@mrao.cam.ac.uk}
\emailAdd{mph@mrao.cam.ac.uk}
\emailAdd{a.n.lasenby@mrao.cam.ac.uk}

\abstract{One of the main challenges of modern cosmology is to 
  investigate the nature of dark energy in our Universe. The
  properties of such a component are normally summarised as a perfect
  fluid with a (potentially) time-dependent equation-of-state
  parameter $w(z)$. We investigate the evolution of this parameter
  with redshift by performing a Bayesian analysis of current
  cosmological observations.  We model the temporal evolution as
  piecewise linear in redshift between `nodes', whose $w$-values and
  redshifts are allowed to vary. The optimal number of nodes is chosen
  by the Bayesian evidence. In this way, we can both determine the
  complexity supported by current data and locate any features present
  in $w(z)$. We compare this node-based reconstruction with some
  previously well-studied parameterisations: the
  Chevallier-Polarski-Linder (CPL), the Jassal-Bagla-Padmanabhan
  (JBP) and the Felice-Nesseris-Tsujikawa (FNT).  
  By comparing the Bayesian evidence for all of these
  models we find an indication towards possible time-dependence in the
  dark energy equation-of-state. It is also worth noting that the CPL
  and JBP models are strongly disfavoured, whilst the FNT is just
  significantly disfavoured, when compared to a simple cosmological constant
  $w=-1$. We find that our node-based reconstruction model
   is slightly disfavoured with respect  to the $\Lambda$CDM model.}
  
  
\begin{document}
  \maketitle
  \flushbottom

\section{Introduction}

Over the past decade, one of the most pressing goals of modern
cosmology has been to explain the accelerated expansion of the
Universe \cite{Riess98,Perlmutter99}.  Considerable observational and
theoretical effort has been focused on understanding this remarkable
phenomenon. It is often postulated that an exotic new source of
stress-energy with negative pressure may be responsible for the cosmic
acceleration: such a component is called {\em dark energy} (DE).

\noindent
The dynamical properties of dark energy are normally summarised as a
perfect fluid with (in general) a time-dependent equation-of-state
parameter $w(z)$, defined as the ratio of its pressure to its energy
density. The simplest proposal, namely a cosmological constant
$\Lambda$, is described by the redshift independent $w=-1$.
Alternative cosmological models that deviate from standard $\Lambda$CDM, but
still lead to an accelerating Universe, include: K-essence,
quintessence and non-minimally coupled scalar fields \cite{Ratra88,Picon00,Vikman05, Urena00}, 
braneworld models \cite{Maartens04}, modified gravity 
\cite{Hu07,Starobinsky07,Appleby07,Capozziello02,Nojiri06}, 
interacting dark energy \cite{Amendola07, Clemson12,Lu12},
anisotropic universes \cite{Akarsu10,Marra12,Valkenburg12}, amongst many others
\cite{Caldwell98, Zlatev99,Cervantes10,Macorra11, Gupta09, Valeria05, Matarrese04, Sahni98}. 
In the absence of a fundamental and well-defined theory of dark
energy, $w(z)$ has been parameterised in a number of different ways,
including: the CPL, JBP and FNT models \cite{CPL1,  CPL2, JBP, FNT}, the Hannestad and Wetterich
parameterisations \cite{Hannestad, Wetterich}, polynomial, logarithmic and oscillatory 
expansions  \cite{Ma11, Sendra11, Zhang11}, Kink models 
\cite{Bassett04}, and quite a few others \cite{Shafieloo09}.  The \emph{a priori} assumption 
of a specific model or the use of particular parameterisations can, however, lead
to misleading results regarding the properties of the dark energy.
Hence, some studies instead perform a direct, model-independent (`free-form')
reconstruction of $w(z)$ from observational data, using, for
instance, a principal component analysis \cite{Huterer03, Zhao08, Serra09,Zhao10,Gong10, Ishida11}, 
maximum entropy techniques  \cite{Zunckel07},
binning $w(z)$ in redshift space \cite{ Putter08,Lazkoz12},
 non-parametric approaches 
\cite{Shafieloo06,Holsclaw10,Holsclaw10b,Shafieloo12,Shafieloo12b,Seikel12}
and several other techniques \cite{Huterer01,Ujjaini04, Linder05,Sahni08,Rubin09, Alam07,
Saini00, Chiba00, Weller02, Alam03, Daly03, Wang04, Wang05,Huterer05, Sahni06}
\\ 

In this paper we explore the possible dynamical behaviour of the dark
energy based on the most minimal \emph{a priori} assumptions. Given
current cosmological observations and using the Bayesian evidence as
an implementation of Occam's razor, we select the preferred shape of
$w(z)$. Our method considers possible deviations from the cosmological
constant by modelling $w(z)$ as a linear interpolation between a set
of `nodes' with varying $w$-values and redshifts (in the most general
case). An advantage of this method is that the number of nodes is
directly chosen by the model Bayesian evidence.  This reconstruction
process is essentially identical to the approach used previously to
recover the preferred shape of the primordial spectrum of curvature
perturbations $P(k)$ \cite{Vazquez12}.  For comparison, we also
consider some existing models that propose a parameterised functional
form for $w(z)$, namely the CPL, JBP and FNT models.  For each model we
compute its evidence and, according to the Jeffreys guidelines, we
select the best model preferred by the data.
 \\
 
The paper is organised as follows: in the next Section we describe the
data sets and cosmological parameters used in the analysis.  We then
describe the form of existing parameterisations used by other authors
and define the reconstruction used in this paper. The resulting
parameter constraints and evidences for each model are then
discussed. Finally, in Section~\ref{sec:conclusion}, based on
Jeffrey's guidelines, we decide which model provides the best
description for current observational data and present our
conclusions.

\section{Analysis}
\label{sec:Pk}

The data-sets considered throughout our analysis include temperature
and polarisation measurements from the 7-year data release of the
Wilkinson Microwave Anisotropy Probe (WMAP; \cite{WMAP}), together with
the 148 GHz measurements from the Atacama Cosmology Telescope (ACT;
\cite{ACT}). In addition to CMB data, we include distance measurements
of 557 Supernovae Ia from the Supernova Cosmology Project Union 2
compilation (SCP; \cite{SCP}).  We also incorporate Baryon Acoustic
Oscillation (BAO; \cite{BAO}) measurements of distance, and baryon
density information from Big Bang Nucleosyntesis (BBN; \cite{BBN}),
and impose a Gaussian prior using measurements of the Hubble parameter
today $H_0$, from the Hubble Space Telescope key project (HST;
\cite{HST}).
\\

We consider purely Gaussian adiabatic scalar perturbations and neglect
tensor contributions.  We assume a flat CDM universe\footnote{The
  possibility of a dynamical dark energy in a curved universe has also
  been considered by, i.e. \cite{Gong11,Hlozek08, Pan10, Wang07}.} described by the
following parameters: $\Omega_{\rm b} h^2$ and $\Omega_{\rm DM} h^2$
are the physical baryon and dark matter densities, respectively,
relative to the critical density ($h$ is the dimensionless Hubble
parameter such that $H_0=100h$ kms$^{-1}$Mpc$^{-1}$), $\theta$ is
$100\times$ the ratio of the sound horizon to angular diameter
distance at last scattering surface, $\tau$ is the optical depth at
reionisation, $A_{\rm s}$ and $n_{\rm s}$ are the amplitude of the
primordial spectrum and the spectral index respectively, measured at
the pivot scale $k_0=0.002$ Mpc$^{-1}$.
Aside from the Sunyaev-Zel'dovich (SZ) amplitude $A_{SZ}$ used by WMAP
analyses, the 148 GHz ACT likelihood incorporates two additional
secondary parameters: the total Poisson power $A_p$ at $l=3000$ and
the amplitude of the clustered power $A_c$.
 To describe the overall shape of the dark energy equation-of-state
 parameter $w(z)$ in our nodal reconstruction, we introduce a set of
 amplitudes $w_{z_i}$ at determined positions $z_i$. The CPL and JBP models each depend upon
 just two parameters: $w_0$ and $w_a$; whereas the FNT parameterisation depends upon four
 parameters:  $w_0$, $w_a$, $\tau$ and $a_{t}$. The assumed flat priors on the
 parameters of each $w(z)$ reconstruction are discussed below.
\\

To carry out the exploration of the parameter space, we input $w(z)$
into a modified version of the {\sc CAMB} code \cite{CAMB}, which
implements a parameterised post-Friedmann (PPF) presciption for the
dark energy perturbations \cite{PPF}. Then, we incorporate into the
{\sc CosmoMC} package \cite{Cosmo} a substantially improved and
fully-parallelized version of the {\it nested sampling} algorithm {\sc
  MultiNest} \cite{Multi1, Multi2}.
The {\sc MultiNest} algorithm increases the sampling efficiency for
calculating the evidence and allows one to obtain posterior samples
even from distributions with multiple modes and/or pronounced
degeneracies between parameters.
The Bayes factor $\mathcal{B}_{ij}$, or equivalently the difference in
log evidences $\ln \mathcal{Z}_i - \ln \mathcal{Z}_j$, provides a
measure of how well model $i$ fits the data compared to model $j$ 
\cite{Saini04,Gordon07, Trotta08, Vazquez12a}. A
suitable guideline for making qualitative conclusions has been laid
out by Jeffreys \cite{Jeffreys}: if $\mathcal{B}_{ij}< 1$ model $i$
should not be favoured over model $j$, $1<\mathcal{B}_{ij}<2.5$ 
constitutes significant evidence, $2.5<\mathcal{B}_{ij}<5$ is strong evidence, while
$\mathcal{B}_{ij}>5$ would be considered decisive.

\subsection{Nodal Reconstruction I}
\label{sec:rec_1}

We first perform the reconstruction of $w(z)$ by parameterising it as
piecewise linear between a set of nodes with variable amplitudes ($w_{z_i}$-values),
but with fixed, equally-spaced redshifts.  Throughout, we bear in mind
that current relevant information, mainly coming from SN Ia, is
encompassed between the present epoch $z_{\rm min} =0$ and $z_{\rm
  max}=2$. At higher redshifts there is no substantial information to
place strong constraints on dark energy, thus beyond $z>2$ we assume $w(z)$
to be constant, with a value equal to that at $z_{\rm max}$. At each
node, we allow variations in amplitudes $w_{z_i}$ with a
conservative prior $w_{z_i}=[-2,0] $.
Our description of $w(z)$ can be summarised as:
\begin{eqnarray} 
w(z) = \left\{ \begin{array}{ll} 

w_{z_{\rm{min}}} 				& \quad	z= 0\\ 
w_{z_i}			      			& \quad	z \in \{z_i\}   \\ 
w_{z_{\rm {max}}}	 			&\quad 	z\ge 2 
\end{array} \right.&& \\  \nonumber  \\
{\rm and \,\,  with \,\,  linear\,\,interpolation \,\,for \quad} 
 0\le z_i<&z&<z_{i+1}\le 2. \nonumber 
\end{eqnarray} 

\begin{figure}[t!]
\begin{center}$
\begin{array}{cc}
(a) \,\,  \mathcal{B}_{1,\Lambda} =-2.19 \pm 0.35 \\ 
\includegraphics[trim = 15mm  -5mm -1mm 5mm, clip, width=6.5cm, height=4.cm]{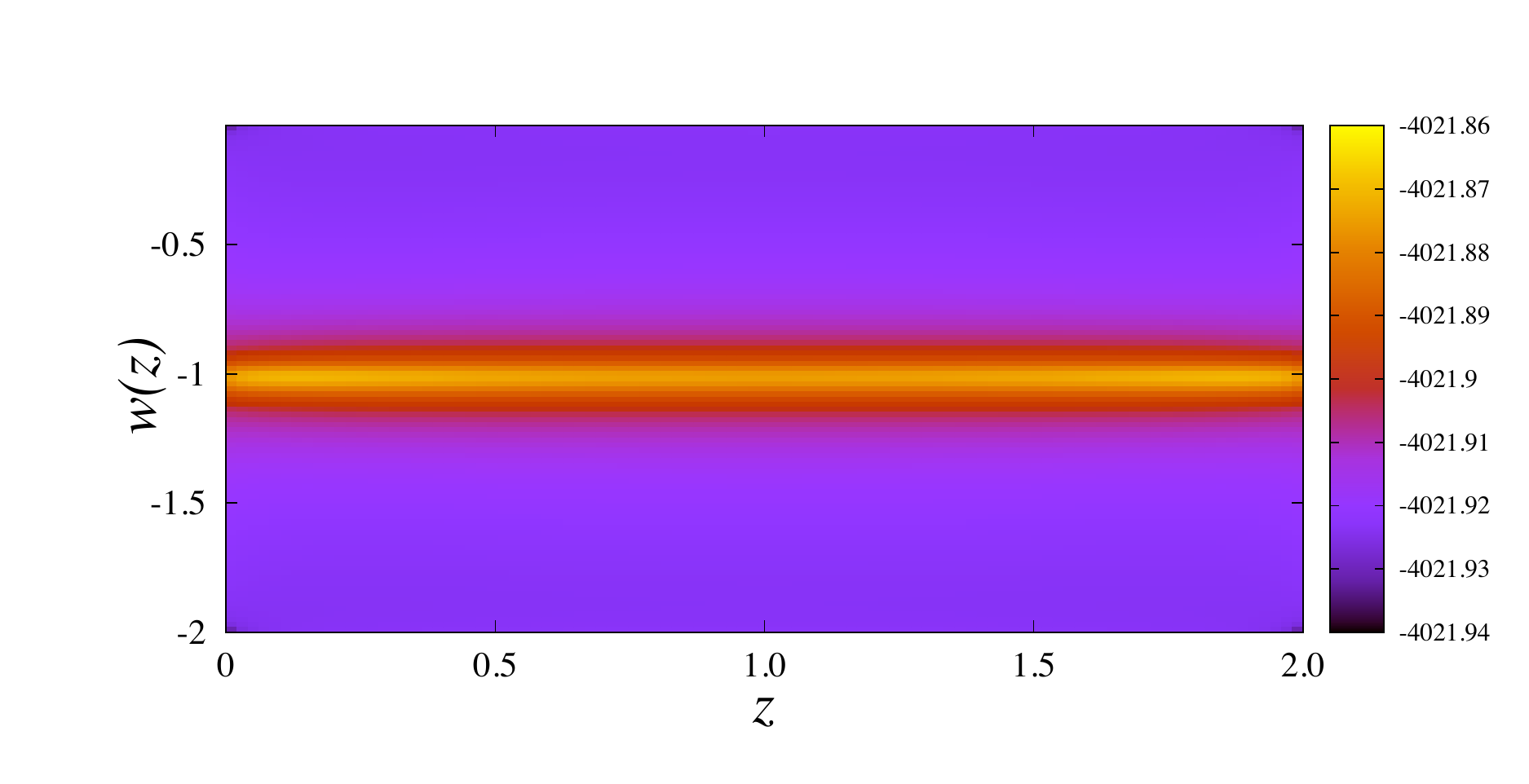}  
\includegraphics[trim = 30mm  75mm 30mm 80mm, clip, width=6cm, height=4cm]{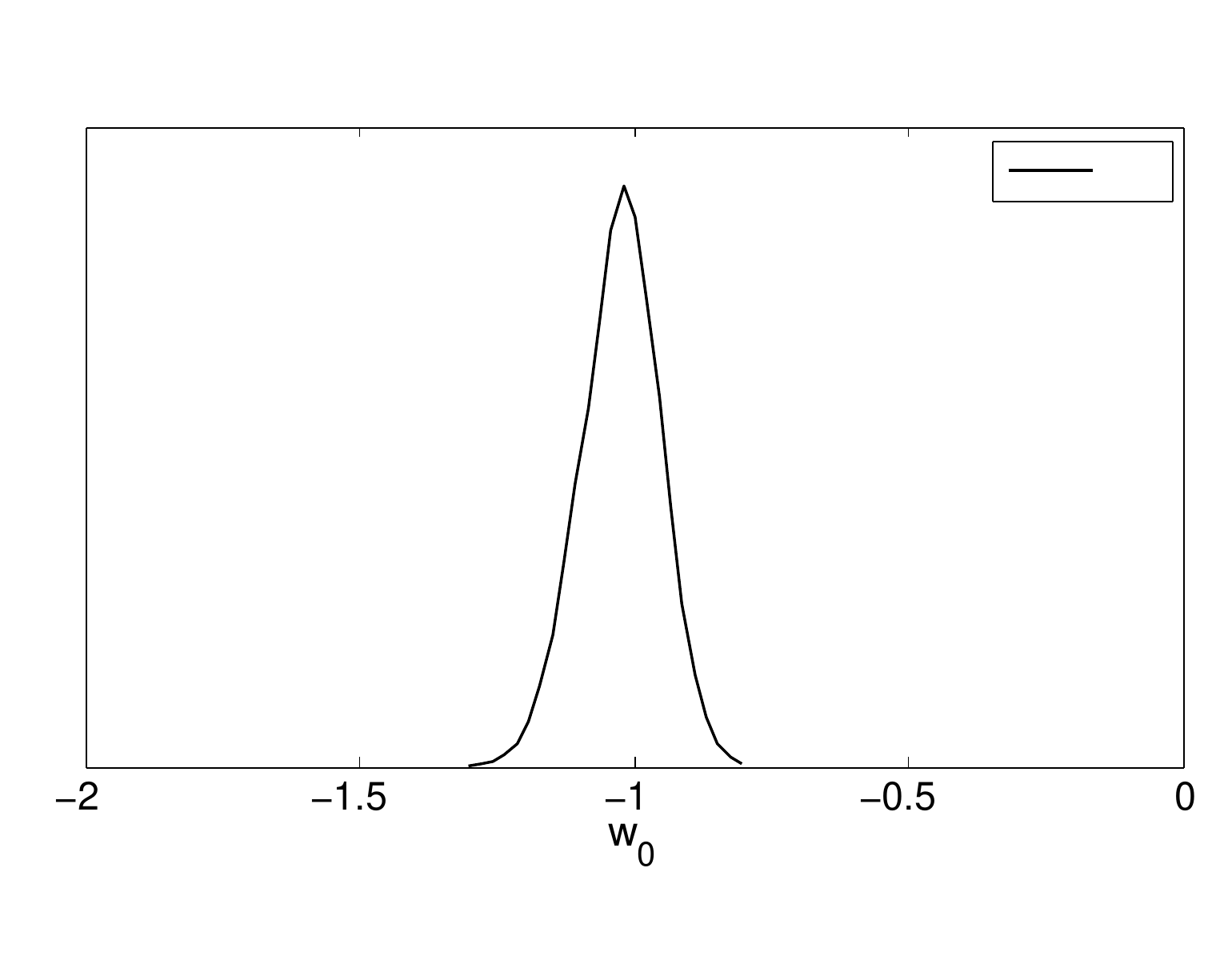}\\
(b) \,\, \mathcal{B}_{2,\Lambda} =-2.34 \pm 0.35 \\ 
\includegraphics[trim = 15mm  -5mm -1mm 5mm, clip, width=6.5cm, height=4.cm]{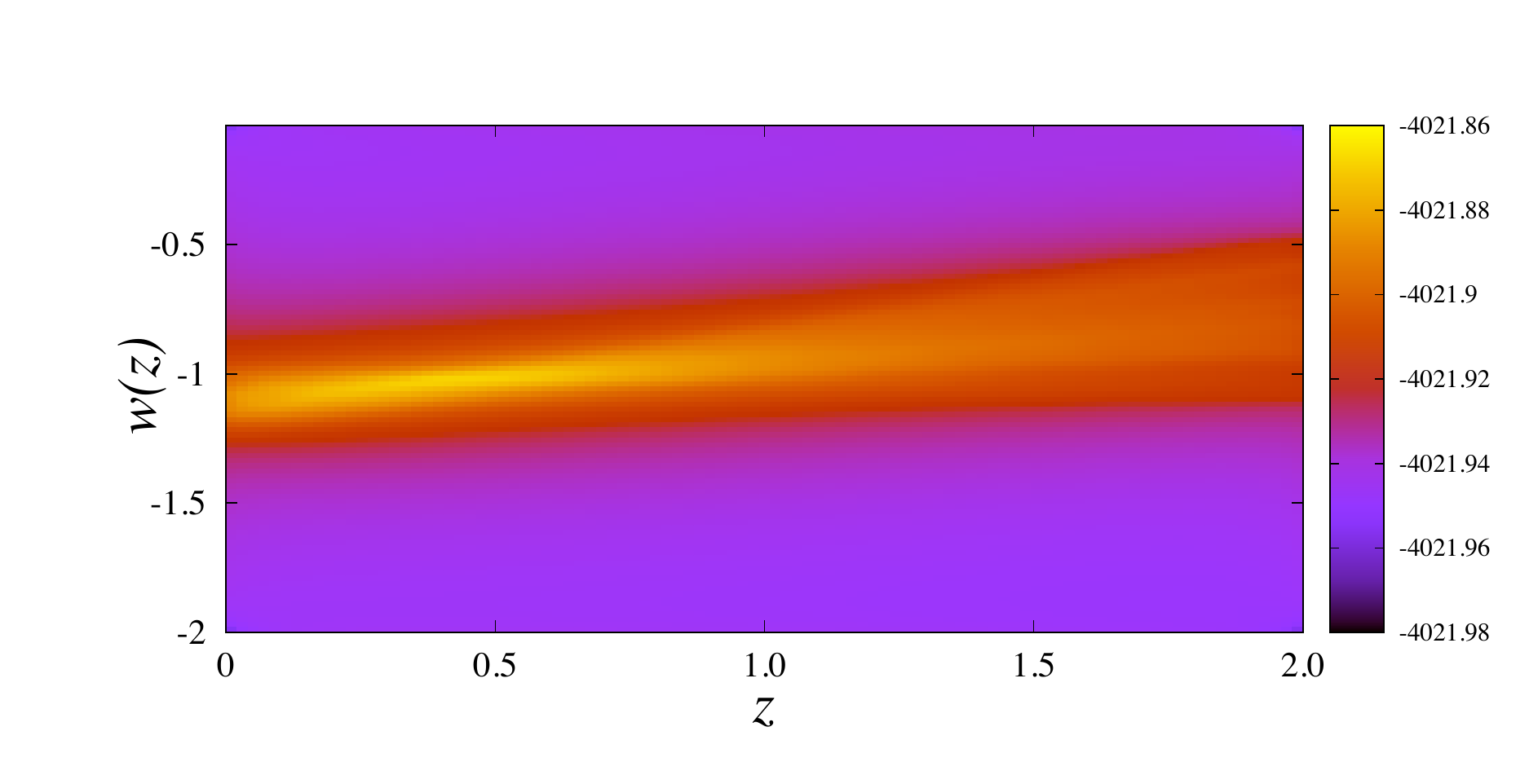}  
\includegraphics[trim = 30mm  75mm 30mm 80mm, clip, width=6cm, height=4cm]{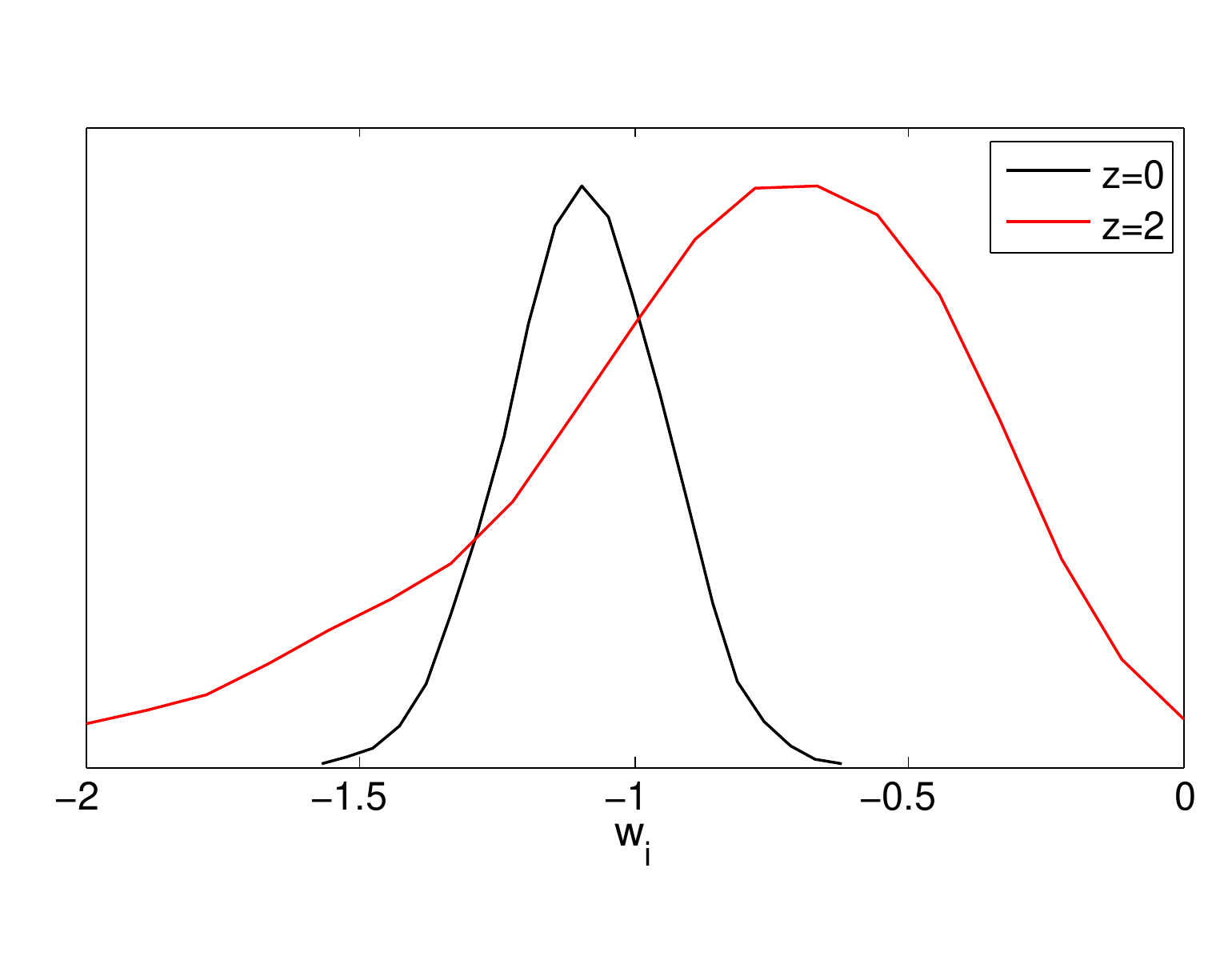}\\
 (c) \,\, \mathcal{B}_{3,\Lambda} =-1.70 \pm 0.35  \\
\includegraphics[trim = 15mm  -5mm -1mm 5mm, clip, width=6.5cm, height=4.cm]{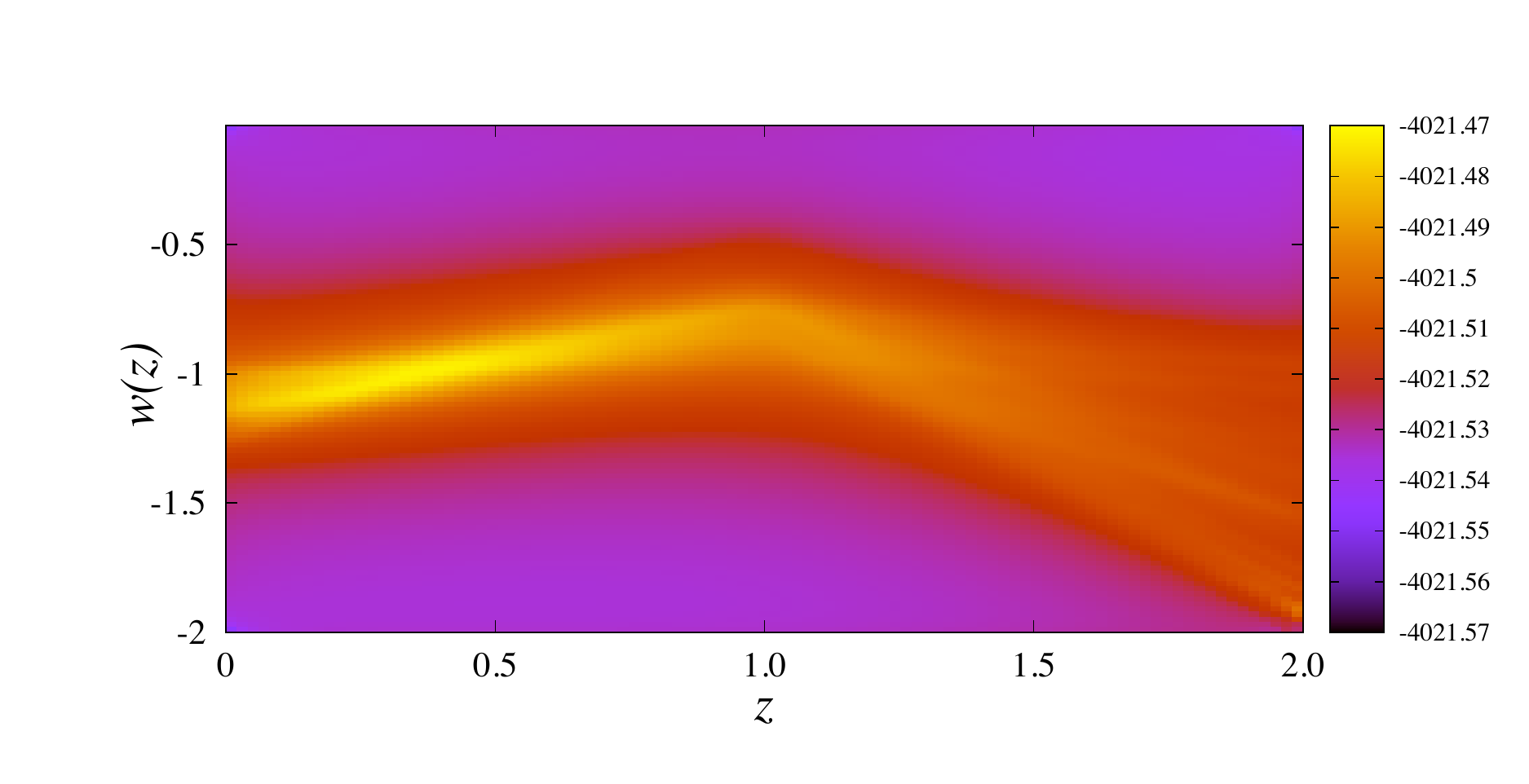}  
\includegraphics[trim = 30mm  75mm 30mm 80mm, clip, width=6cm, height=4cm]{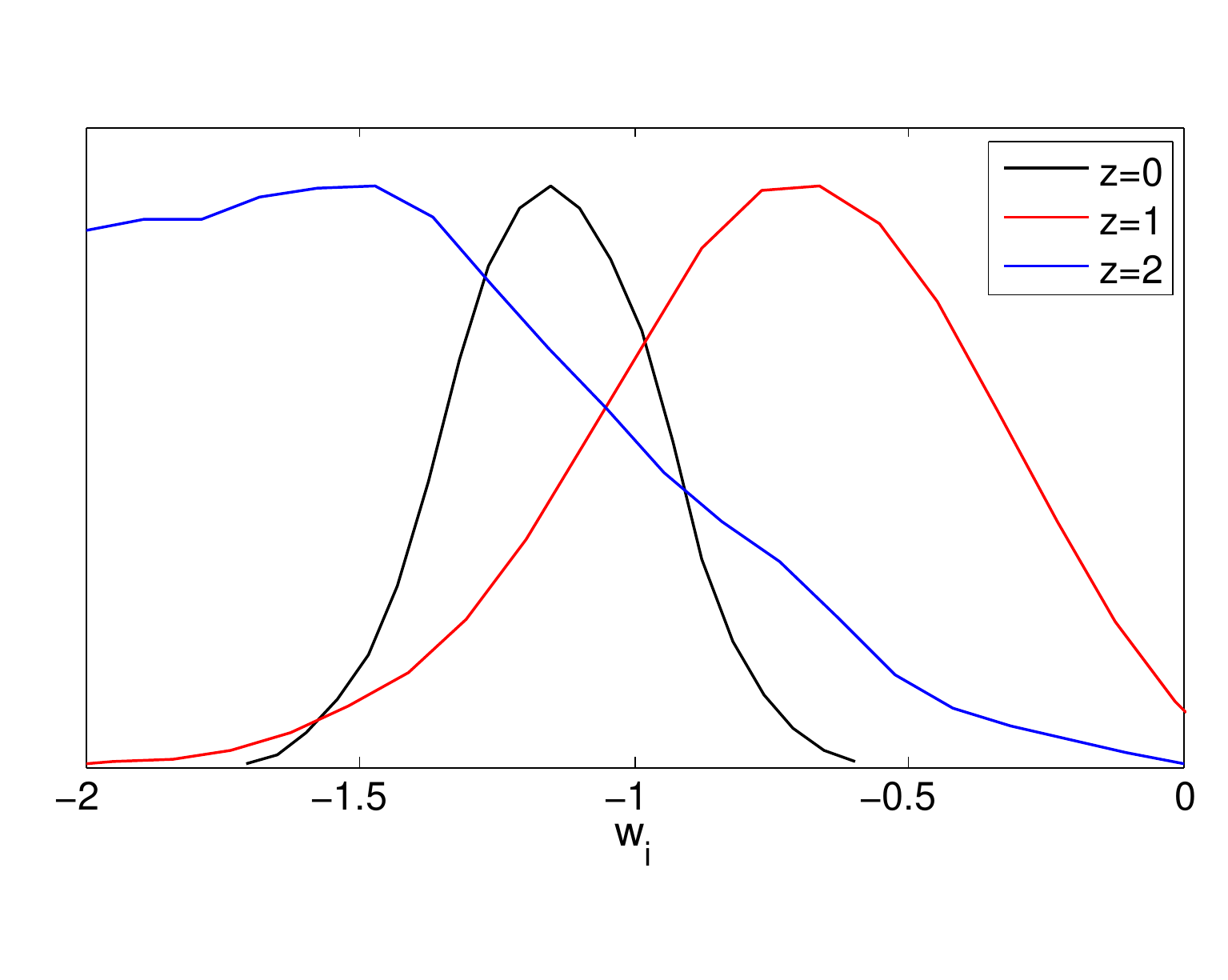}\\
 (d) \,\, \mathcal{B}_{4,\Lambda} =-1.57 \pm 0.35  \\ 
\includegraphics[trim = 15mm  -5mm -1mm 5mm, clip, width=6.5cm, height=4.cm]{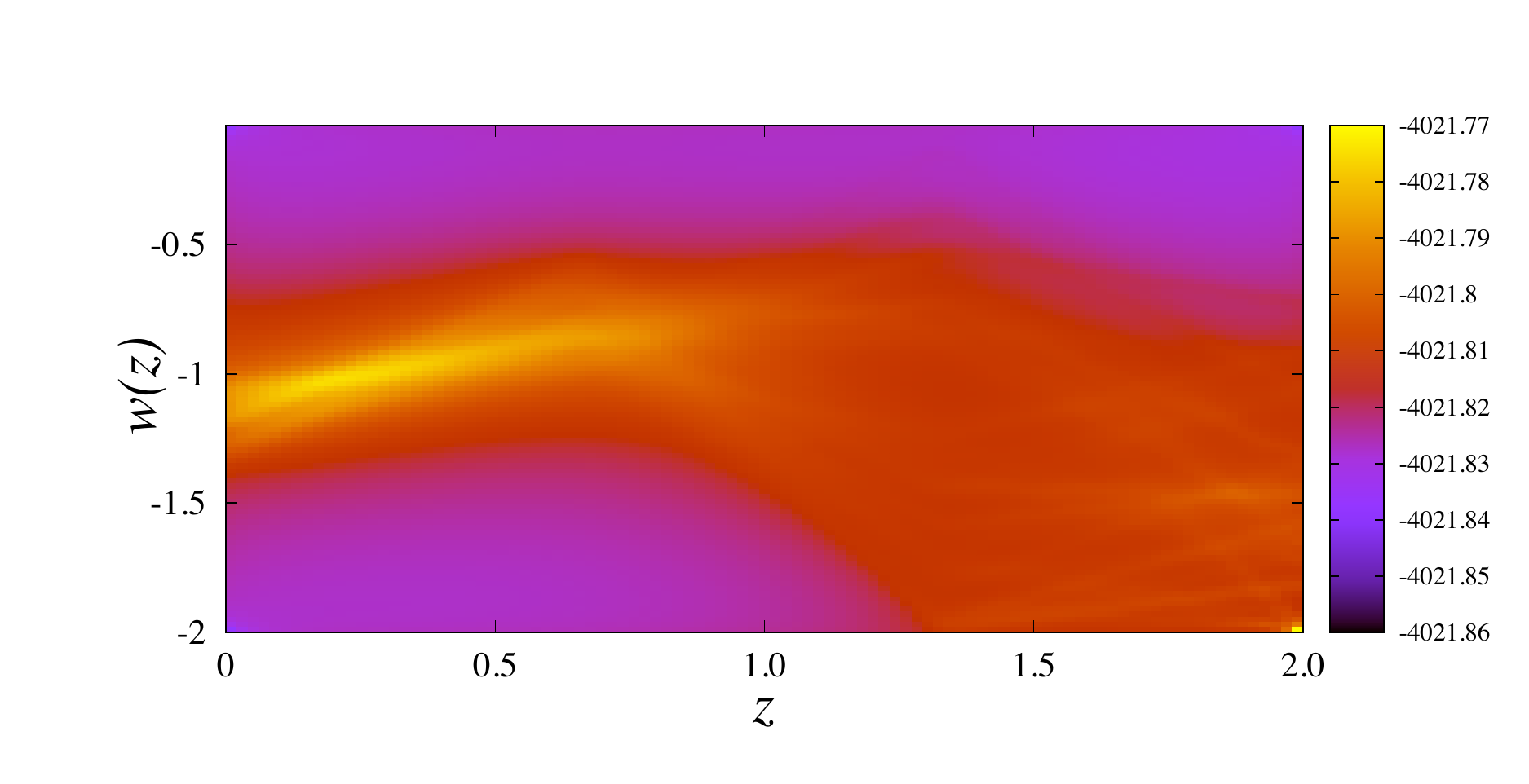}
\includegraphics[trim =  30mm  75mm 30mm 80mm, clip, width=6cm, height=4cm]{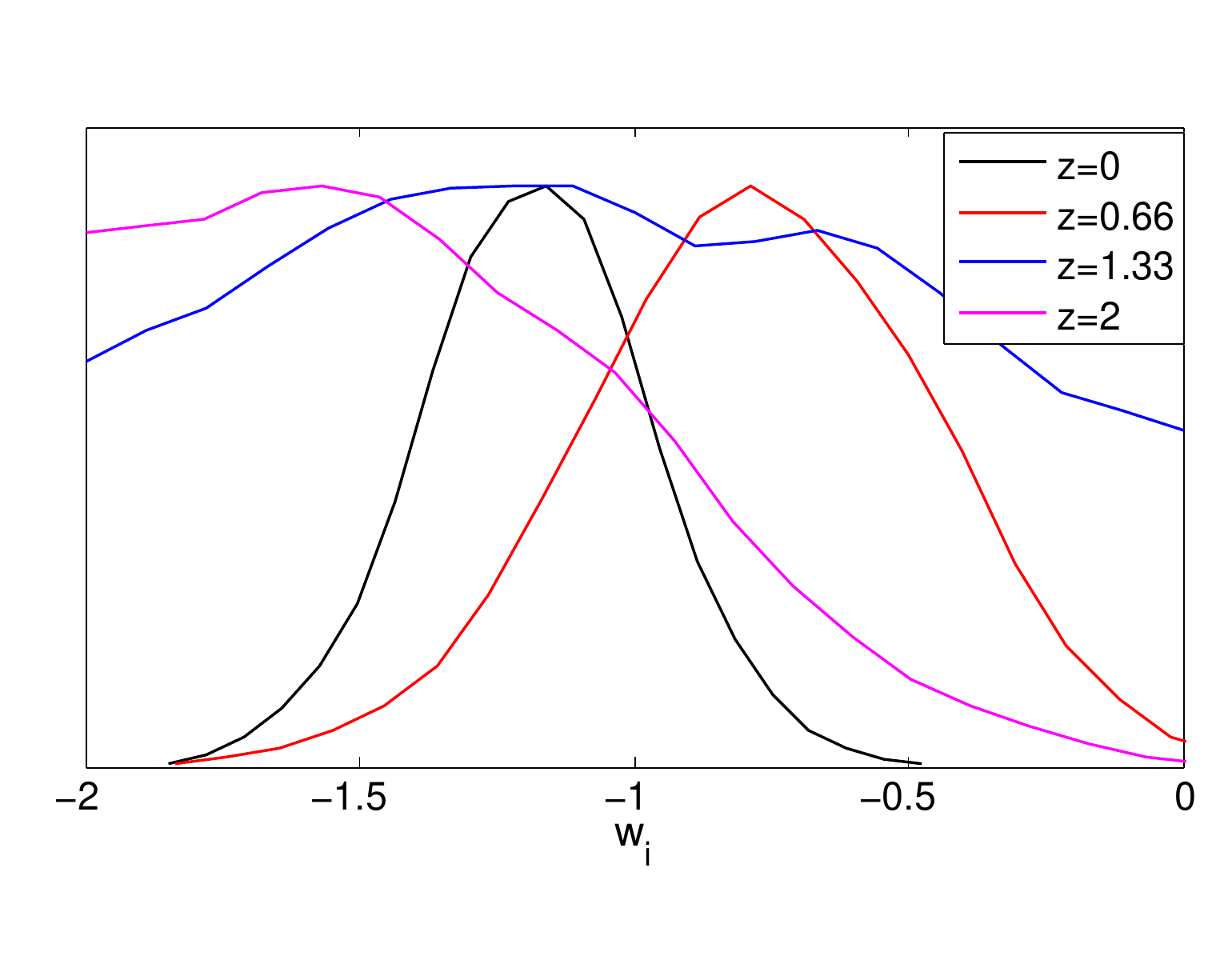}\\
\end{array}$
\end{center}
\caption{Left: Reconstruction of the dark energy equation-of-state
  parameter modelled as piecewise linear between nodes that may vary
  in amplitude $w_i$ but are fixed in redshift $z$, showing
  the mean amplitude values and their corresponding $1\sigma$
  error bands.  The colour-code shows $\ln ({\rm
    likelihood})$, where lighter regions represents an improved fit.
  Right: 1D marginalised posterior distribution of the amplitudes
  $w_i$ at each $z$-node (shown in the right-top corner), in each
  reconstruction.  The top label in each panel denotes the associated
  Bayes factor respect to the $\Lambda$CDM model.}
\label{fig:recons_1}
\end{figure}

\noindent
While the use of linear interpolation between nodes may seem crude, we
have shown in a previous work \citep{Vazquez12} that the use of
smoothed interpolation functions, such as cubic splines, can lead to
significant spurious features in the reconstruction, thus leading to
poor fits to observational data and also  unrepresentative errors.
\\

We perform all of our model comparisons with respect to the simplest
explanation of dark energy, namely a cosmological constant, which is
specified by a redshift-independent $w=-1$.  First, we consider
deviations of the $\Lambda$CDM model by letting the equation-of-state
parameter vary only in amplitude: $w(z)=w_0= $ constant (see Figure
\ref{fig:recons_1}(a)).  The incorporation of two or more parameters,
as in models (b) and (c) respectively, allows us to test the dark
energy time-evolution. Figure \ref{fig:recons_1} also includes the 1D
marginalised posterior distribution for the corresponding amplitude at
each node and for each reconstruction. In the top label of each model
we have included the Bayes factor compared to the $\Lambda$CDM model.

In model (b), we notice the overall shape of $w(z)$ includes a
slight positive tilt and a narrow waist located at $z\sim 0.3$. It is
also observed that at the present epoch $w(z=0) \lesssim -1$ is slightly
favoured, while at higher redshifts $ w(z) \gtrsim -1$ is preferred, hence, the
reconstructed $w(z)$ exhibit a crossing of the line $w=-1$.
The crossing of the phantom divide line $w=-1$ (PDL), plays  a key  role
in identifying the correct dark energy model \cite{Zhang09}. If future surveys confirm its 
existence, single scalar field theories (with minimal assumptions) might be in serious problems as they
cannot reproduce this essential feature, and therefore alternative models should
be considered, e.g. scalar-tensor theories \cite{Boisseau00,Deffayet10}, braneworld models 
\cite{Chimento06, Sahni03}, $f(R)$ gravity \cite{Hu07,Appleby07,Starobinsky07,Motohashi11}.
To continue with our reconstruction process, we then place a third point
(c) midway between the two existing nodes in (b).  This model mimics a
running behaviour by allowing slight variations in the interpolated
slopes between the three nodes.  The freedom in its shape, together
with the very weak constraints at high redshifts, lead to a $w(z)$
with slight negative slope at early times, in contrast to model
(b). Furthermore the presence of a small bump in the resulting $w(z)$
at $z\sim 1$ (see Figure \ref{fig:recons_1} (c)) might point to some
weak departure from the cosmological constant $w=-1$.

\noindent
We can continue this process of adding more nodes but always using the
Bayesian evidence to penalise any unnecessary inclusion of model
parameters. The inclusion of a fourth stage with $z$-space split into
three equally spaced regions is given by model (d).  At low redshifts
the shape of the equation of state is well constrained with tight
error bands on each node, whereas at high redshifts the error bands
again indicate the lack of sufficient data to provide strong
constraints.  Notice also the increased error bands due to the
addition of further nodes and (anti-)correlations
created between them: for instance, the posterior distribution of the
amplitude $w_{z_i}$ at $z=0$ is broadened as the number of nodes is
increased.  At this stage, the evidence has flattened off, and so
it seems reasonable to stop adding parameters in the
reconstruction process at this point.
The constraints on the $w_{z_i}$ amplitudes used on each reconstruction
are given by (for two-tailed distributions 68\% C.L.  are shown, whilst for 
one-tailed distributions the upper 95\% C.L.):
\begin{eqnarray*}
(a)&&  w_0=-1.02\pm 0.07, \\
(b)&&  w_{z=0}=-1.09\pm 0.14,   \quad   w_{z>2}=-0.83\pm 0.39,\\
(c)&&  w_{z=0}=-1.14\pm 0.17,   \quad w_{z=1}=-0.73\pm 0.33,    \qquad w_{z>2}<-0.65, \\
(d)&& w_{z=0}=-1.18\pm 0.20,   \quad w_{z=0.66}=-0.78\pm 0.30,   \quad w_{z=1.33}=1.03\pm 0.53, \quad w_{z>2}<-0.62.
\end{eqnarray*}

The models used in the reconstruction of $w(z)$ are assessed according
to the Jeffreys guideline.  The Bayes factor between the $\Lambda$CDM
model and the one-node model $\mathcal{B}_{1,\Lambda}=-2.19 \pm 0.35$
points out that $w(z) = w_0$ (where  $w_0$ is a free constant), is strongly disfavoured when compared to
the cosmological constant, similarly, when two independent nodes are
used $\mathcal{B}_{2,\Lambda} =-2.34 \pm 0.35$.  Thus,
parameterisations that contain one or two parameters are not able to
provide an adequate description of the behaviour of $w(z)$, and hence
are strongly disfavoured by current observations. The addition of
nodes in the third and fourth stage provides more flexibility in
the shape of the reconstructed $w(z)$.  Thus, the evidence for these
models shows an improvement, compared to the first and second models,
indicating the possible presence of some features in the time
evolution of the equation-of-state parameter.  Nonetheless, when they
are compared to $\Lambda$CDM, they are still marginally disfavoured: $
\mathcal{B}_{3,\Lambda} =-1.70 \pm 0.35$ and $\mathcal{B}_{4,\Lambda}
=-1.57 \pm 0.35$.

\subsection{Nodal reconstruction II}
\label{sec:rec_2}

We previously reconstructed $w(z)$ by placing nodes at particular
fixed positions in $z$-space. However, to localise features, we now
extend the analysis by also allowing the $z$-position of each node to
move freely. In particular, we again fix two $z$-nodes at sufficiently
separated positions $z_{\rm min}=0$ and $z_{\rm max}=2$, but now place
inside additional `nodes' with the freedom to move around in both
position $z_i$ and amplitude $w_{z_i}$.  This method has the advantage
that we do not have to specify the number and location of nodes
describing $w(z)$; indeed, the form of any deviation from flat $w(z)$
can be mimicked through a change in the amplitudes and/or positions of
the internal nodes. Also, the reduced number of internal nodes avoids
the creation of wiggles due to high (anti-)correlation between nodes,
which might lead to a misleading shape for $w(z)$. We use 
the same priors for the amplitudes $w_{z_i}=[-2,0]$ as we adopted in
Section \ref{sec:rec_1}.
%
Hence, for this type of nodal-reconstruction the equation of
state is described by
\begin{eqnarray} 
w(z) = \left\{ \begin{array}{ll} 

w_{z_{\rm min}} 				& \quad	z= 0\\ 
w_{z_1}			      		& \quad	0< z_i<z_{i+1}< 2   \\ 
w_{z_{\rm max}}	 			&\quad 	z\ge 2 
\end{array} \right.&& \\ \nonumber \\ 
{\rm and\,\, with\,\,  linear \,\, interpolation\,\, for\,\,}\,\,  0\le &z_1<z_{i+1}&\le 2. \nonumber 
\end{eqnarray} 

%


\begin{figure}
\begin{center}$
\begin{array}{cc}
(z_1)\,\,  $$ \mathcal{B}_{z_1,\Lambda} =-1.27 \pm 0.35 \\ 
 \includegraphics[trim =  15mm  -5mm -1mm 5mm, clip, width=6.5cm, height=4.2cm]{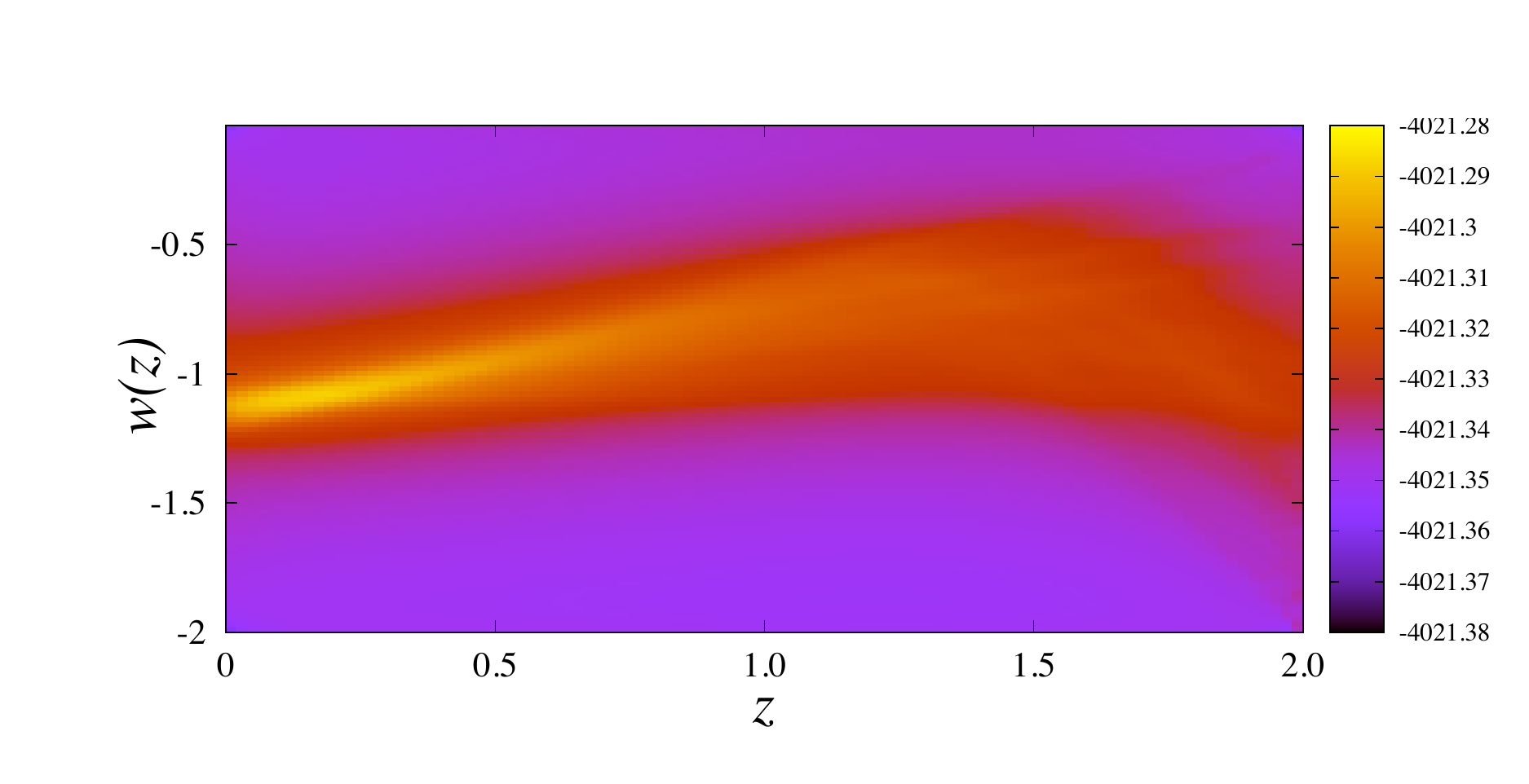} 
\includegraphics[trim =25mm  70mm 30mm 82mm, clip, width=5.cm, height=4.2cm]{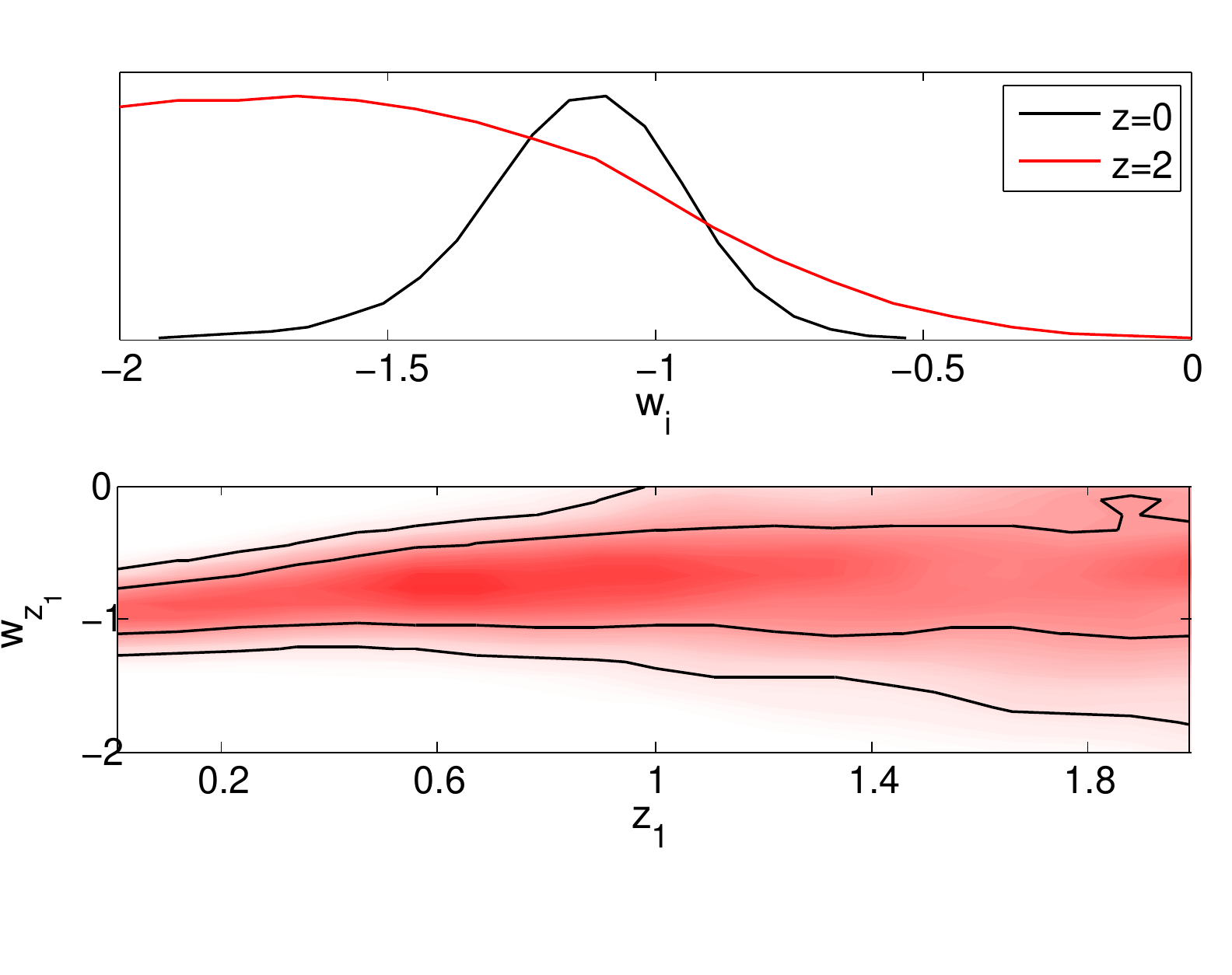}\\
(z_2)\,\, $$\mathcal{B}_{z_2,\Lambda} =-0.81 \pm 0.35$$ \\
 \includegraphics[trim = 15mm  -5mm -1mm 5mm, clip, width=6.5cm, height=4.2cm]{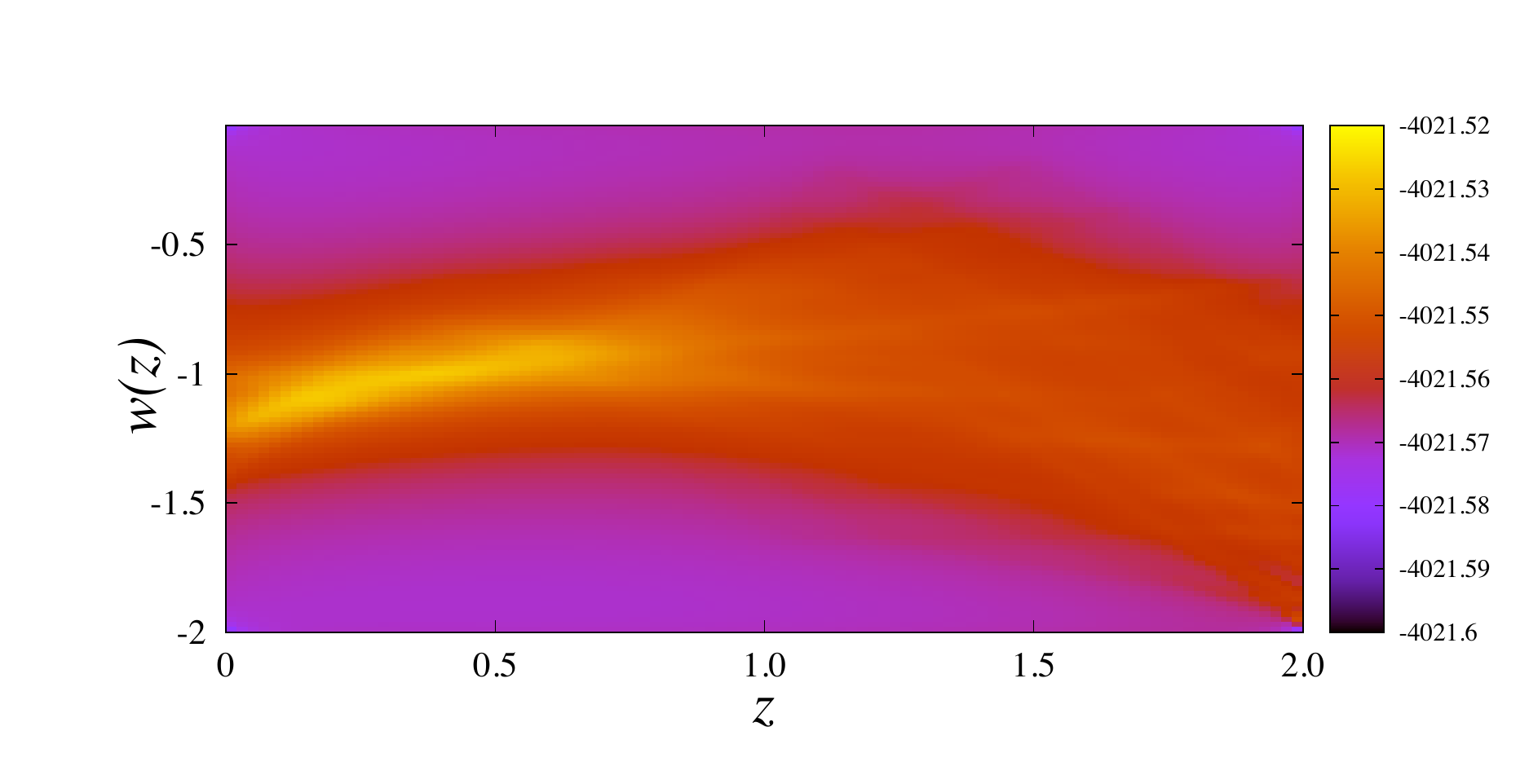}
\includegraphics[trim =25mm  70mm 30mm 82mm, clip, width=5cm, height=4.2cm]{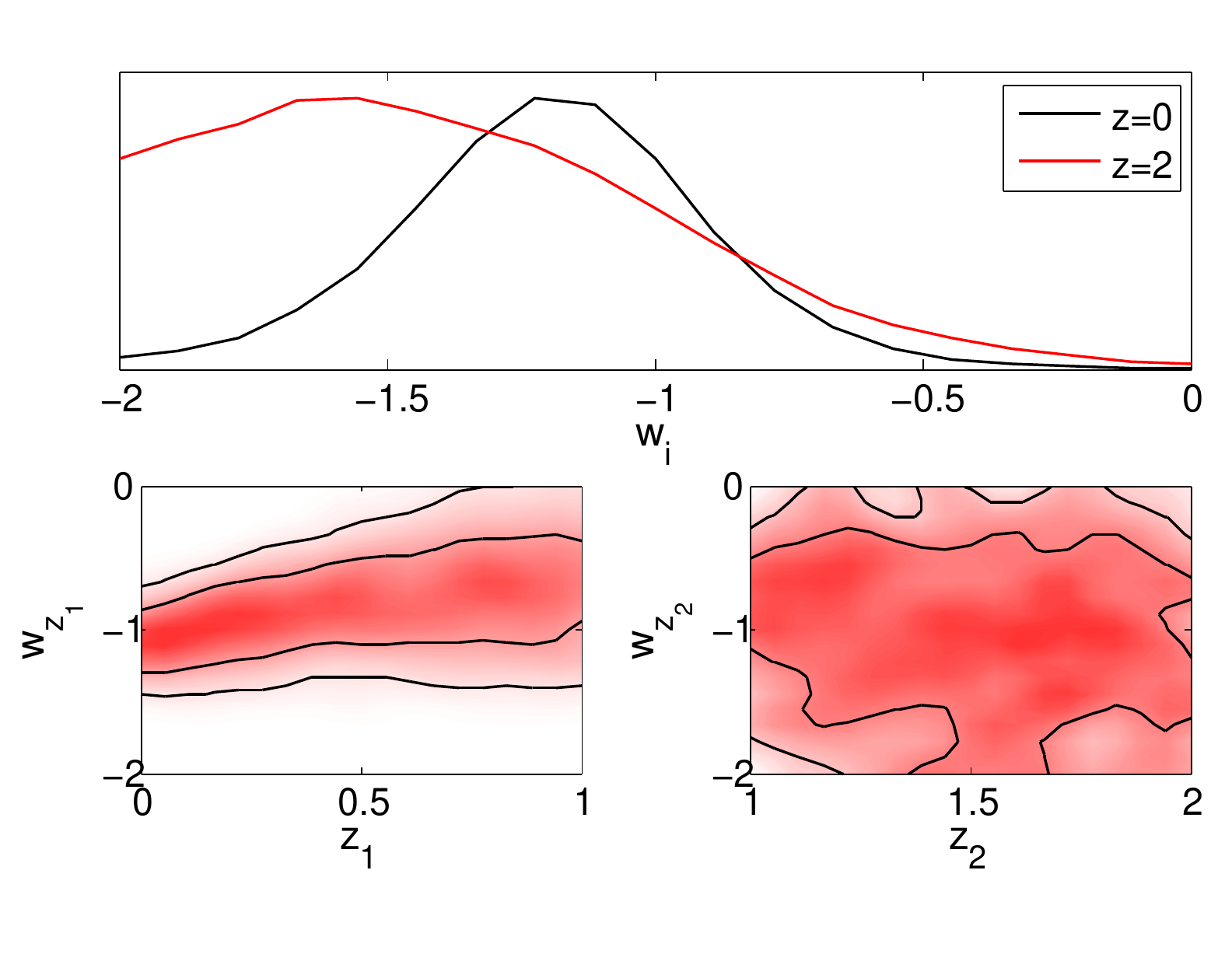}\\
(z_3)\,\, $$\mathcal{B}_{z_3,\Lambda} =-0.95 \pm 0.35$$\\
 \includegraphics[trim = 15mm  -5mm -1mm 5mm, clip, width=6.5cm, height=4.2cm]{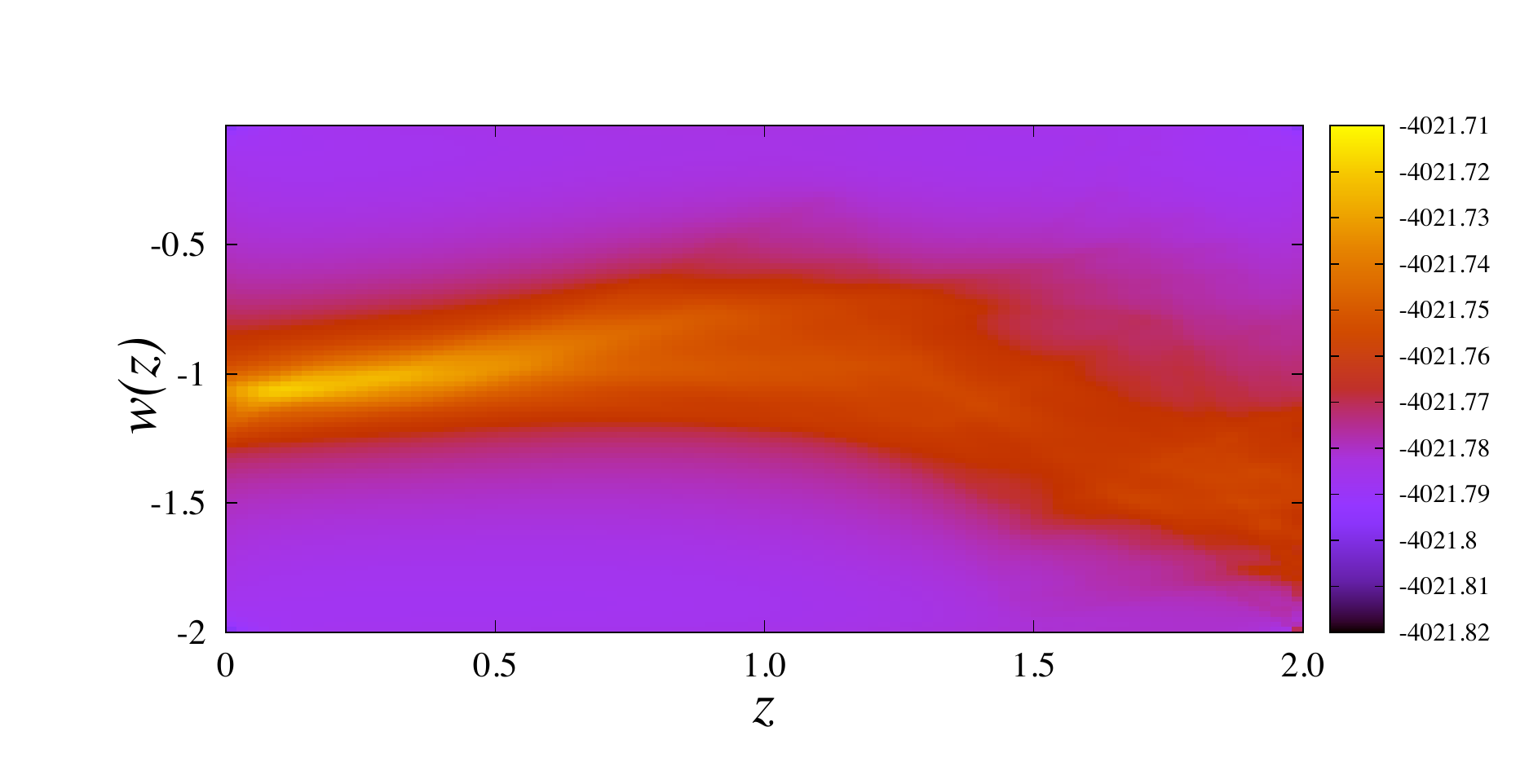}
\includegraphics[trim =25mm  70mm 30mm 82mm, clip, width=5cm, height=4.2cm]{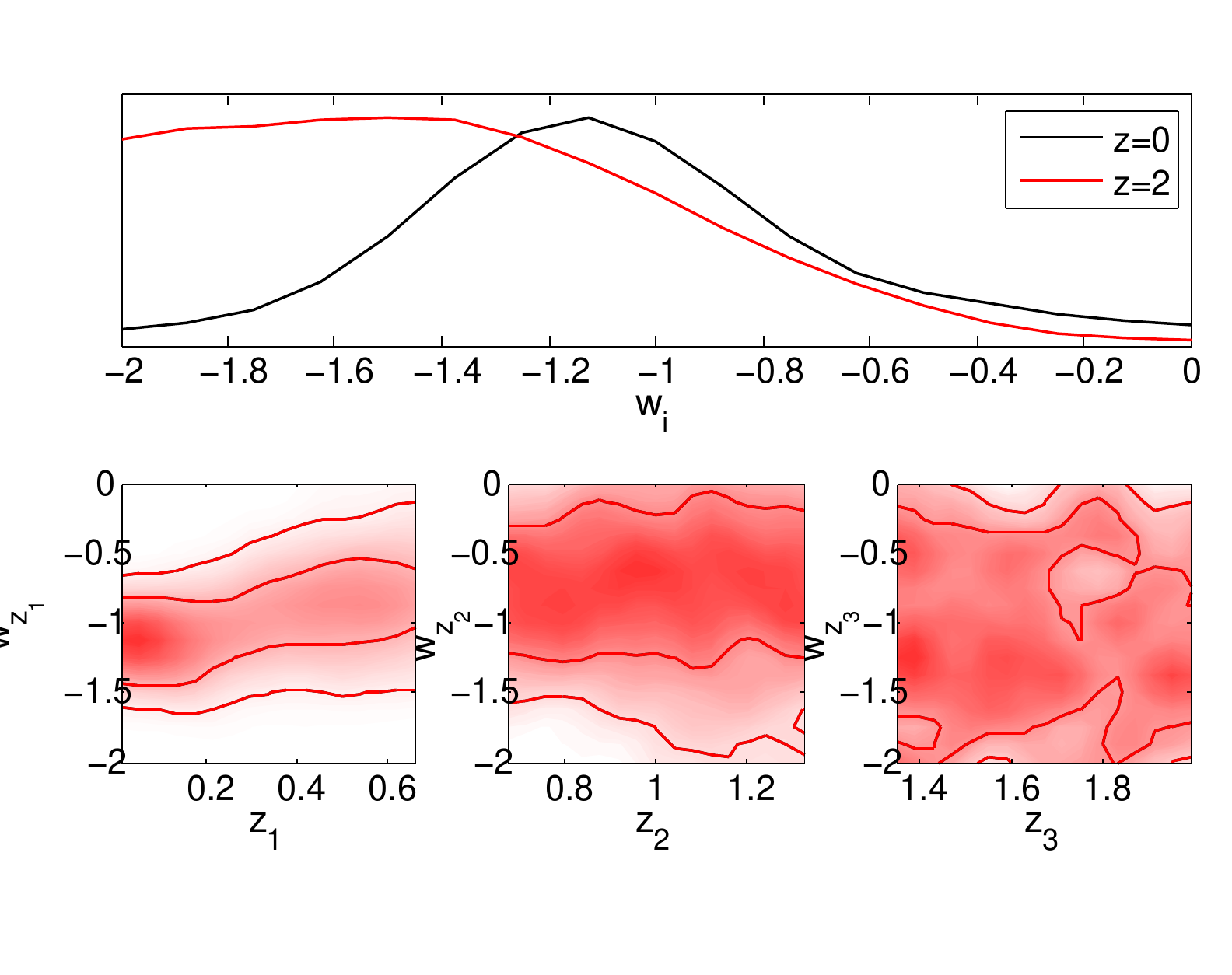}

\end{array}$
\end{center}
\caption{Left: Reconstruction of the dark energy equation-of-state parameter
  $w(z)$ using one-internal-node (top) and two-internal $z$-nodes
  (bottom) that move freely in both amplitude $w_i$ and redshift
  $z_i$. Right: corresponds to the 1D and 2D marginalised posterior
  distribution of the amplitudes and $z$-node positions in each
  reconstruction.  The colour-code indicates the $\ln ({\rm
    Likelihood})$, where lighter regions represents an improved fit,
  and the top label in each panel denotes the associated Bayes factor
  with respect to the $\Lambda$CDM model. }
\label{fig:mov_z}
\end{figure}

\noindent
Figure \ref{fig:mov_z} illustrates the reconstruction of $w(z)$ from
the mean posterior estimates for each node, with $1\sigma$ error bands
on the amplitudes (left). Also plotted are the 1D and 2D marginalised
posterior distributions on the parameters used to describe $w(z)$
(right). The reconstructed shape for the two-internal-node model (middle panel)
resembles the form obtained in Figure \ref{fig:recons_1}(c), but now
with a turn-over shifted to earlier times. A similar turn-over has been found 
using principal component analysis by \cite{Serra09,Gong10}.
The narrow waist at $z\sim0.3$ is also noticeable, where the SNe constraints seem
to be tightest.
For the one and three-internal-nodes case (top and bottom panel of 
Figure \ref{fig:mov_z}), we observe $w(z)$ has essentially  the 
same behaviour as in the two-internal-node model, being the preferred model.
Finally, a common feature throughout all the reconstructed equation of state $w(z)$ is observed:
the presence of the crossing PDL within the range $0<z<0.5$.
The constraints on the $w_{z_i}$ amplitudes used on each reconstruction
are given by (for two-tailed distributions 68\% C.L.  are shown, whilst for 
one-tailed distributions the upper 95\% C.L.):
\begin{eqnarray*}
(z_1)&& w_{z=0}=-1.14\pm 0.18,   \quad w_{0<z<2}>-1.39\pm 0.35,  \quad w_{z>2}<-0.70, \\
(z_2)&& w_{z=0}=-1.18\pm 0.26,   \quad w_{0<z<1}=-0.83\pm 0.29,   \quad w_{1<z<2}=1.02\pm 0.52, \quad w_{z>2}<-0.63,\\
(z_3)&& w_{z=0}=-1.07\pm 0.36 ,  \quad w_{0<z<0.66}=-0.98\pm 0.29,   \quad w_{0.66<z<1.33}=-0.84\pm 0.47, \\
&&    w_{1.33<z<2}=-1.02\pm 0.55,   \quad w_{z>2}<0.63. 
\end{eqnarray*}

The  similar shape of the  three models are in good agreement with their Bayes factor:
 $\mathcal{B}_{z_2,z_1} =+0.46 \pm 0.35$,  $\mathcal{B}_{z_3,z_2} =-0.14 \pm 0.35$.
According to the Jeffreys guideline, even though  the two internal-node  model  contains more parameters, 
it  is  significantly preferred  over the models  with one and  two  fixed-nodes, i.e. 
$\mathcal{B}_{z_2,2} =+1.53 \pm 0.35$. However, when compared  to the  cosmological  constant model 
the Bayes  factor is too  small  to draw  any decisive conclusions: $\mathcal{B}_{z_2,\Lambda} =-0.81 \pm 0.35$.
Thus we conclude that the internal-node models might be considered as viable models to characterise the 
dark energy dynamics.
As seen in Figure \ref{fig:mov_z}, the Bayesian evidence has 
reached a plateau and thus we cease the addition of further nodes.

\subsection{CPL and JBP  parameterisations}
\label{subsec:CPL}

\begin{figure}
\begin{center}$
\begin{array}{cc}
 ({\rm CPL})\,\, $$ \mathcal{B}_{{\rm CPL},\Lambda} =-2.84 \pm 0.35 \\ 
 \includegraphics[trim =  15mm  -5mm -1mm 5mm, clip, width=6cm, height=4.2cm]{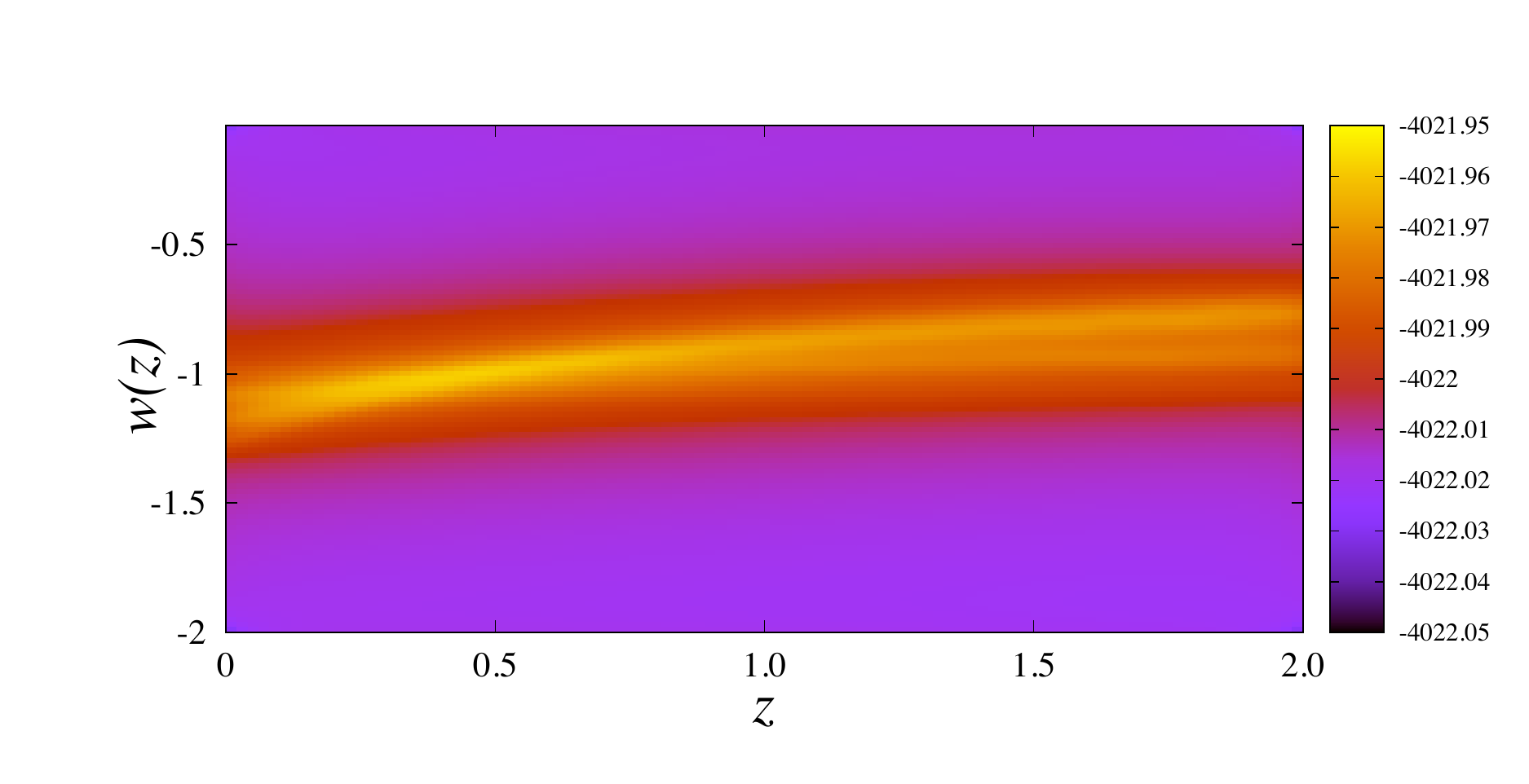} 
\includegraphics[trim =20mm  70mm 30mm 90mm, clip, width=5.5cm, height=4.0cm]{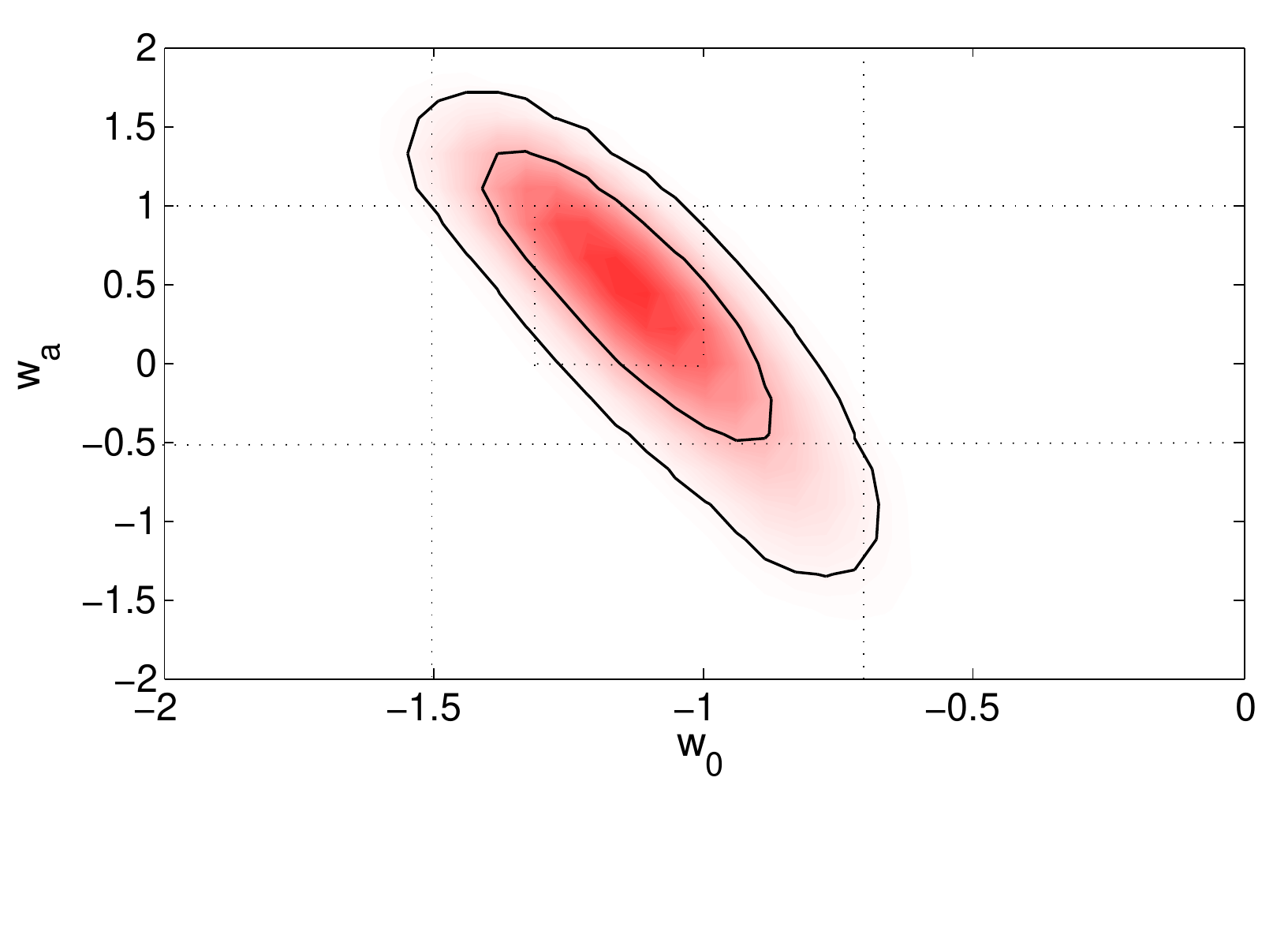}\\
 ({\rm JBP})\,\, $$ \mathcal{B}_{{\rm JBP},\Lambda} =-2.82 \pm 0.35 \\ 
 \includegraphics[trim =  15mm  -5mm -1mm 5mm, clip, width=6cm, height=4.2cm]{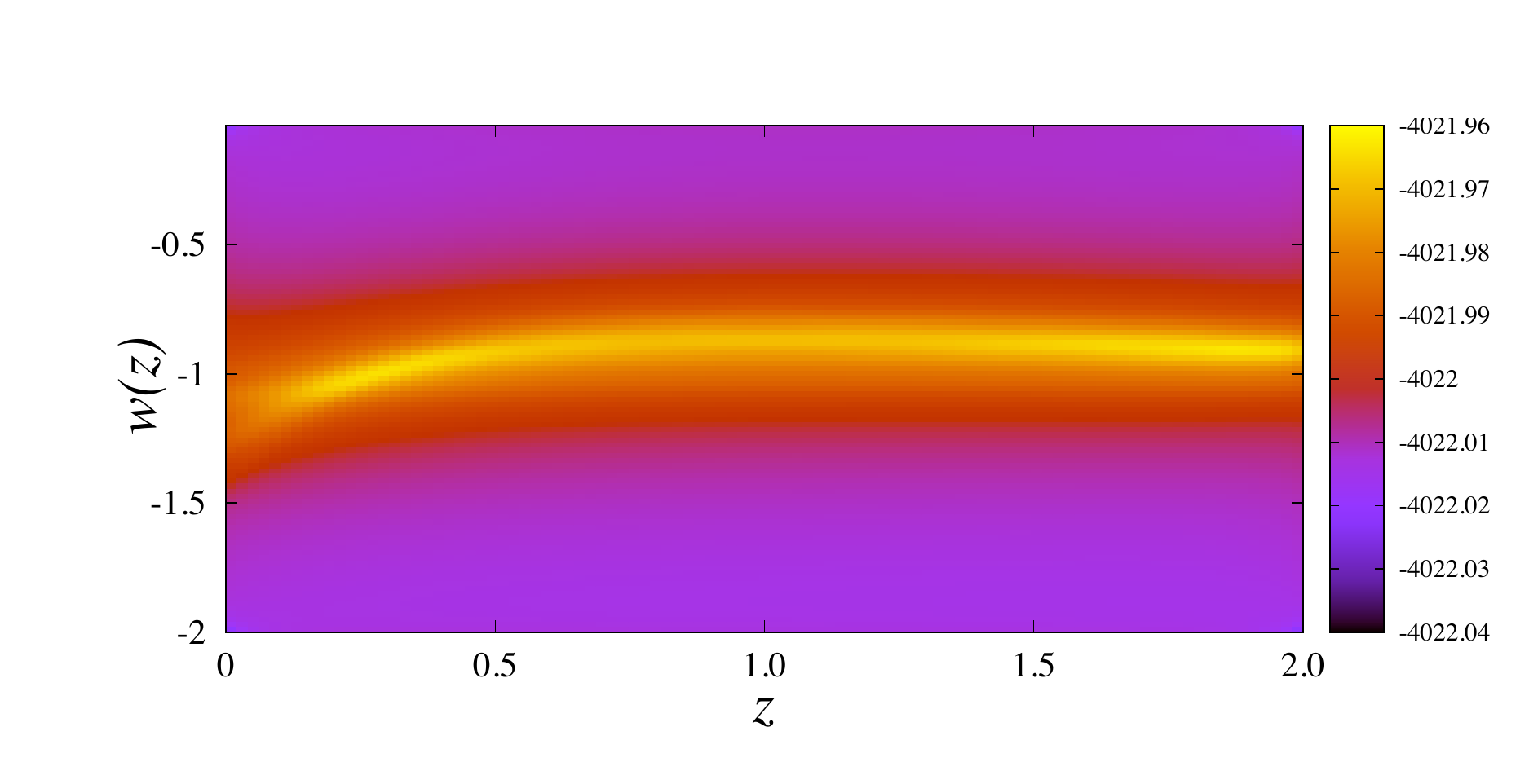} 
\includegraphics[trim =20mm  70mm 30mm 90mm, clip, width=5.5cm, height=4.0cm]{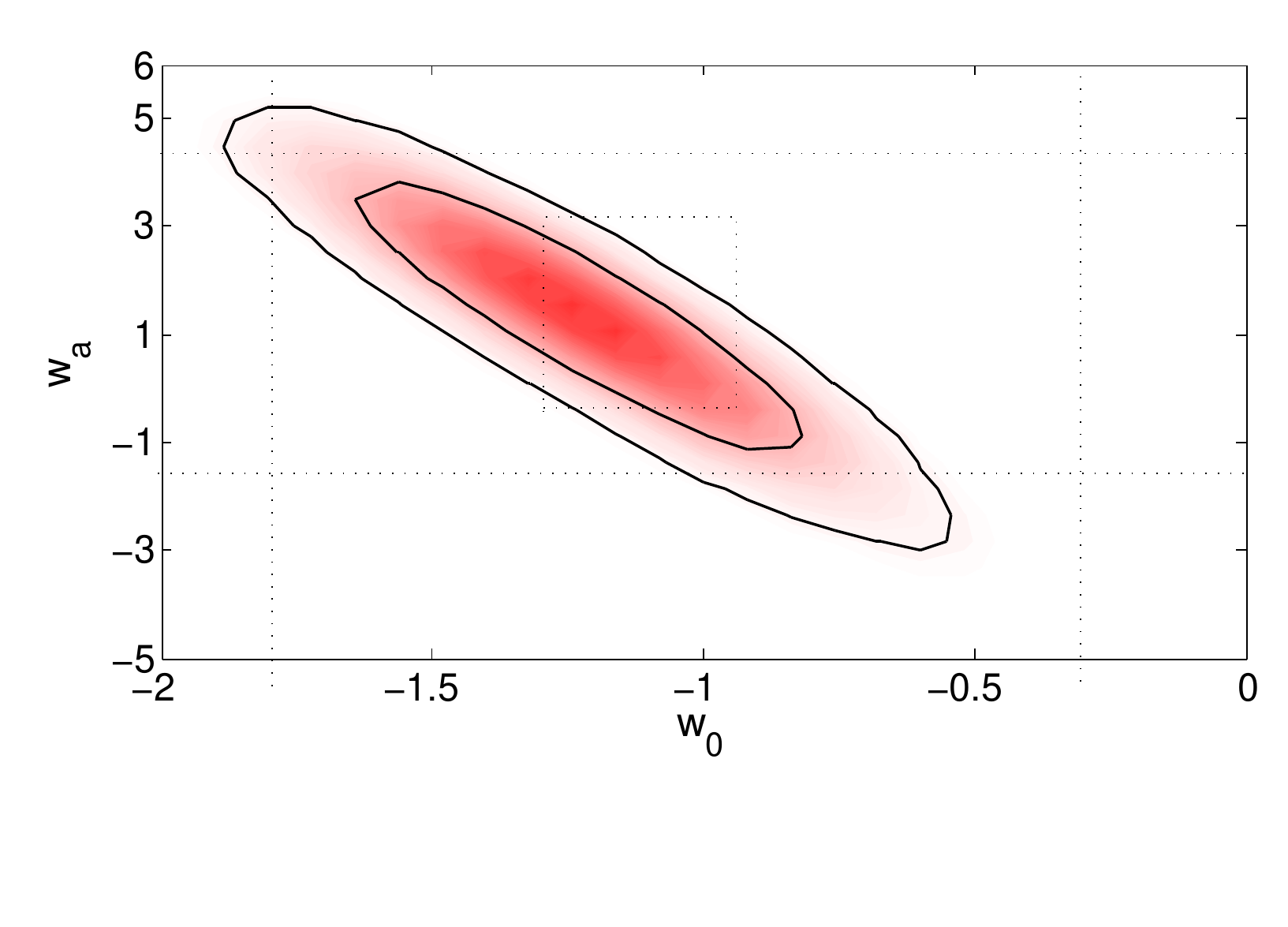}
\end{array}$
\end{center}
\caption{Reconstruction of the dark energy equation of state $w(z)$
  assuming the Chevallier-Polarski-Linder (top) and the
  Jassal-Bagla-Padmanabhan parameterisation (bottom), along with their
  corresponding 2D constraints with $1\sigma$ and $2\sigma$ confidence
  contours (right panel).  The colour-code indicates the $\ln ({\rm
    Likelihood})$, where lighter regions represents an improved fit;
  the top label in each panel denotes the associated Bayes factor with
  respect to the $\Lambda$CDM model. Dotted lines indicate the priors
  choice.}
\label{fig:cpl}
\end{figure}

In this section we examine some existing parameterised models for
$w(z)$ and compare these to our nodal reconstructions.
In particular, we consider the simple parameterised description
introduced by Chevallier-Polarski-Linder (CPL;~\cite{CPL1, CPL2}), that
has the functional form:
\begin{equation} \label{eq:CPL}
w(z)= w_0 + w_a\frac{z}{1+z},
\end{equation}
where the parameters $w_0$ and $w_a$ are real numbers such that at the
present epoch $w|_{z=0}=w_0$ and $dw/dz|_{z=0}=-w_a$, and as we go back in
time $w(z \gg 1)\sim w_0 + w_a$.
Thus, we limit the CPL parameters by the flat priors $w_0=[-2,0]$
and $w_a=[-3,2]$.
\\

\noindent
We also consider the parameterisation suggested by
Jassal-Bagla-Padmanabhan (JBP;~\cite{JBP}):
\begin{equation}
w(z)= w_0 + w_a\frac{z}{(1+z)^2}.
\end{equation}
In this model, the parameter $w_0$ determines the properties of $w(z)$
at both low and high redshifts: $w(z=0)=w_0$ and $w(z \gg 1)\sim w_0$.
To explore the parameter space we consider the following flat priors
on the JBP parameters: $w_0=[-2,0]$ and $w_a=[-6,6]$. 
\\

Figure \ref{fig:cpl} shows 2D joint constraints, with $1\sigma$ and
$2\sigma$ confidence contours, for the parameters used to describe the
CPL and JBP models, and the resulting shape of $w(z)$ corresponding to
the mean posterior estimates of $w_0$ and $w_a$.  In each panel we
have included the Bayes factor compared to the $\Lambda$CDM model.
Both of the models are in good agreement with a simple
cosmological constant. The current constraints for the CPL and JBP parameters 
are essentially as we expected: 

\begin{eqnarray*}
({\rm CPL}) &&w_0=-1.11\pm0.17, \quad w_a=0.34\pm 0.60, \\
({\rm JBP}) &&  w_0= -1.21\pm 0.26, \quad w_a=1.28\pm1.62. 
\end{eqnarray*}

Given that the CPL and JBP parametererisations depend upon just two
parameters, they seem to not posses enough freedom to capture local
features of $w(z)$, i.e. the CPL model does not exhibit 
a turn-over, 
see Figure \ref{fig:cpl}. This is reflected in the large
difference in the Bayesian evidence for this model compared to that of the cosmological
constant: $ \mathcal{B}_{{\rm CPL},\Lambda} =-2.84 \pm 0.35$ and
$ \mathcal{B}_{{\rm JBP},\Lambda} =-2.82 \pm 0.35 $.  In fact, the CPL
equation of state looks similar to that obtained in Figure
\ref{fig:recons_1} (b), confirming our results.  An important point to
emphasise is that, for the chosen priors, $\mathcal{B}_{{\rm CPL},z_2}
=-2.03 \pm 0.35$ and $\mathcal{B}_{{\rm JBP},z_2} =-2.01 \pm 0.35$,
indicating that both models are strongly disfavoured in comparison to
the internal-node reconstruction, shown in Figure \ref{fig:mov_z}.
\\

To illustrate the robustness of the model to small variations of the
prior range, we compute the Bayesian evidence using different sets of priors, shown in Table
\ref{tab:priors}; the prior ranges are illustrated with dotted lines
in Figure \ref{fig:cpl}.  The reader will observe that even though the priors,
in the first three choices, have been shrunk to within the region of
the $2\sigma$ contours, the Bayes factor still disfavours significantly
both the CPL and JBP parameterisations compared to the cosmological
constant and  the two-internal-node  reconstruction.
With respect to the extremely small prior (last row of Table
\ref{tab:priors}), we notice that the JBP model does not contain the
cosmological constant $w_0=-1$. Its Bayes factor compared
to the $\Lambda$CDM model $ \mathcal{B}_{{\rm JBP},\Lambda} =-0.54
\pm 0.35$, shows that models with $w(z=0)\lesssim-1.1 $ might provide a good
description for the current state of the Universe.

\begin{table}
\begin{center}
\caption{Robustness
of the CPL and JBP models over small variations of the prior range.
The associated Bayes factor in each  model is  compared with  respect
to the $\Lambda$CDM model.}

\begin{tabular}{cccc} 
\cline{1-4}\noalign{\smallskip}
\vspace{0.2cm}
Prior   &   $ \mathcal{B}_{{\rm CPL},\Lambda}$ \qquad&\qquad Prior  & $ \mathcal{B}_{{\rm JBP},\Lambda}$ \\
$w_0,w_a$&\qquad&\qquad$w_0,w_a$&\\
 \hline
\vspace{0.2cm}
[-1.5,-0.7], [-3,2]		& $-1.84 \pm 0.35$\qquad\qquad&	 [-1.8,-0.6], [-6,6] &	 $-2.35\pm 0.35$	\\
\vspace{0.2cm}
[2,0], [-0.5,1]		& $-2.11 \pm 0.35$\qquad\qquad&	 [-2,0], [-1,4] &	  $-1.82 \pm 0.35$	\\
\vspace{0.2cm}
[-1.5,-0.7], [-0.5,1]	& $-1.39 \pm 0.35$\qquad\qquad&	 [-1.8,-0.6], [-1,4] &	 $-1.51 \pm 0.35$	\\
\vspace{0.2cm}
[-1.3,-1], [0,1]		&  $-0.26 \pm 0.35$ \qquad\qquad&	 [-1.4,-1.1], [0,3] &	 $-0.54 \pm 0.35$	\\
\hline
 
\end{tabular}
\label{tab:priors}
\end{center}
\end{table}

\subsection{FNT parameterisation}

We have observed that two-parameter functions
are not, in general, sufficient to recover the evolution of the dark energy $w(z)$,
obtained previously in the reconstruction process.  
As an alternative to the CPL and JBP functional form, we consider a more general parameterisation
introduced by Felice-Nesseris-Tsujikawa (FNT, \cite{FNT}), which allows fast transitions for the dark energy
equation of state:

\begin{equation}
w(a)=w_a+(w_0-w_a)\frac{a^{1/\tau}[1-(a/a_{t})^{1/\tau}]}{1-a_t^{-1/\tau}},
\end{equation}

\noindent
where $a=1/(1+z)$, $a_t>0$ and $\tau>0$.
The parameter $w_0$ determines the $w(a)$ properties
at present time $w_0=w(a=1)$, whereas $w_a$ the asymptotic past $w_a=w(a\ll1)$. 
In this model, the equation of state $w(a)$ has an extremum
at $a_{*}=a_t/2^{\tau}$ with value

\begin{equation}\label{eq:fnt}
w(a_*)=w_p+\frac{1}{4}\frac{(w_0-w_a)a_t^{1/\tau}}{1-a_t^{-1/\tau}}.
\end{equation}

\noindent

\noindent
Based on the assumptions given by \cite{FNT},
we explore the cosmological parameter-space using the following flat priors:
 $w_0=[-2,0]$, $w_a=[-2,0]$, $a_t=[0,1]$ and $\tau=[0,1]$,
 using a full Monte-Carlo exploration. We leave the analysis
 of the robustness of this model under small variations on the priors, for a future work.
\\

\begin{figure}
\begin{center}$
\begin{array}{cc}
 ({\rm FNT})\,\, $$ \mathcal{B}_{{\rm FNT},\Lambda} =-1.68 \pm 0.35 \\ 
 \includegraphics[trim =  15mm  -5mm -1mm 5mm, clip, width=6cm, height=4.2cm]{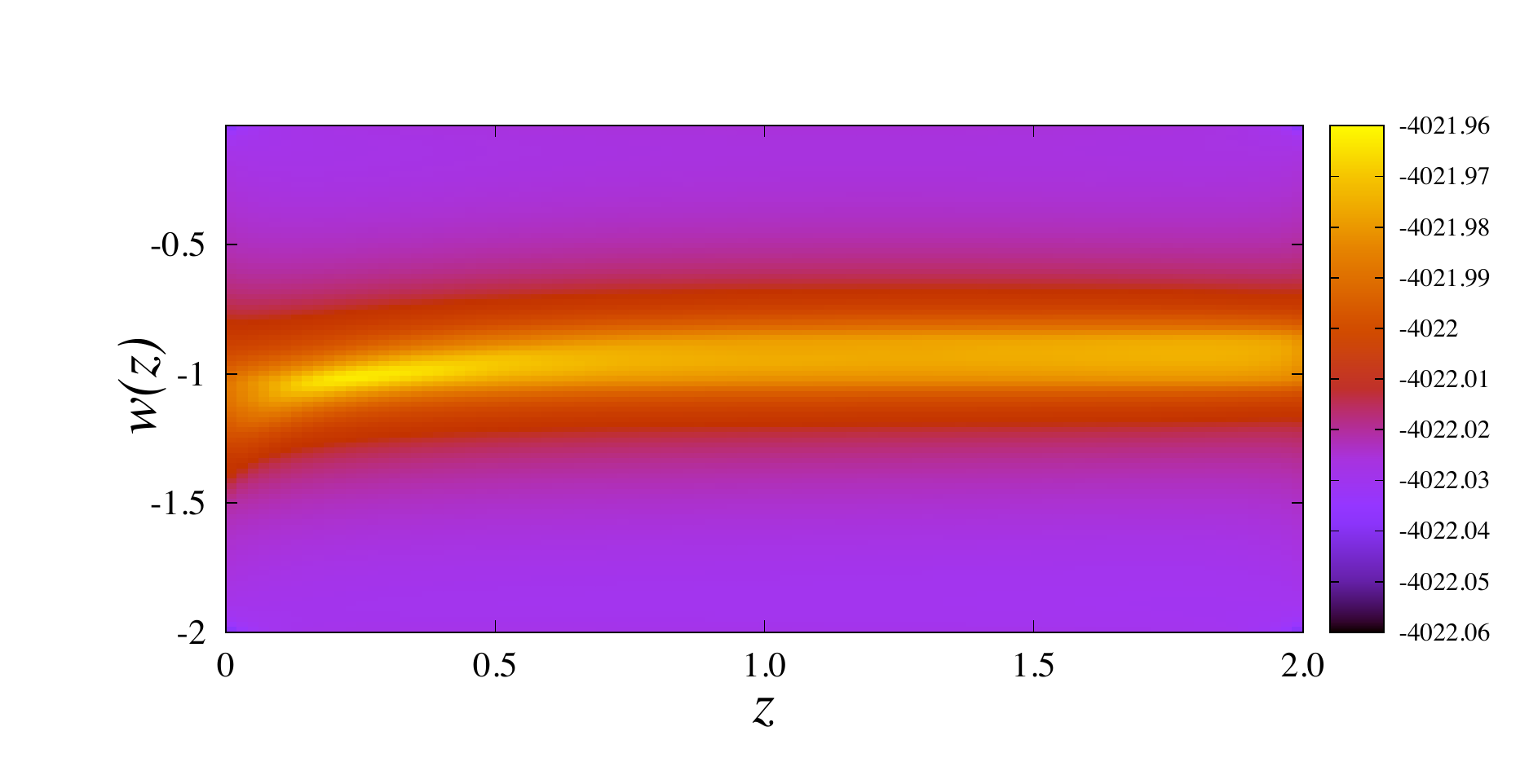} 
\includegraphics[trim =20mm  40mm 3mm 40mm, clip, width=5.5cm, height=4.5cm]{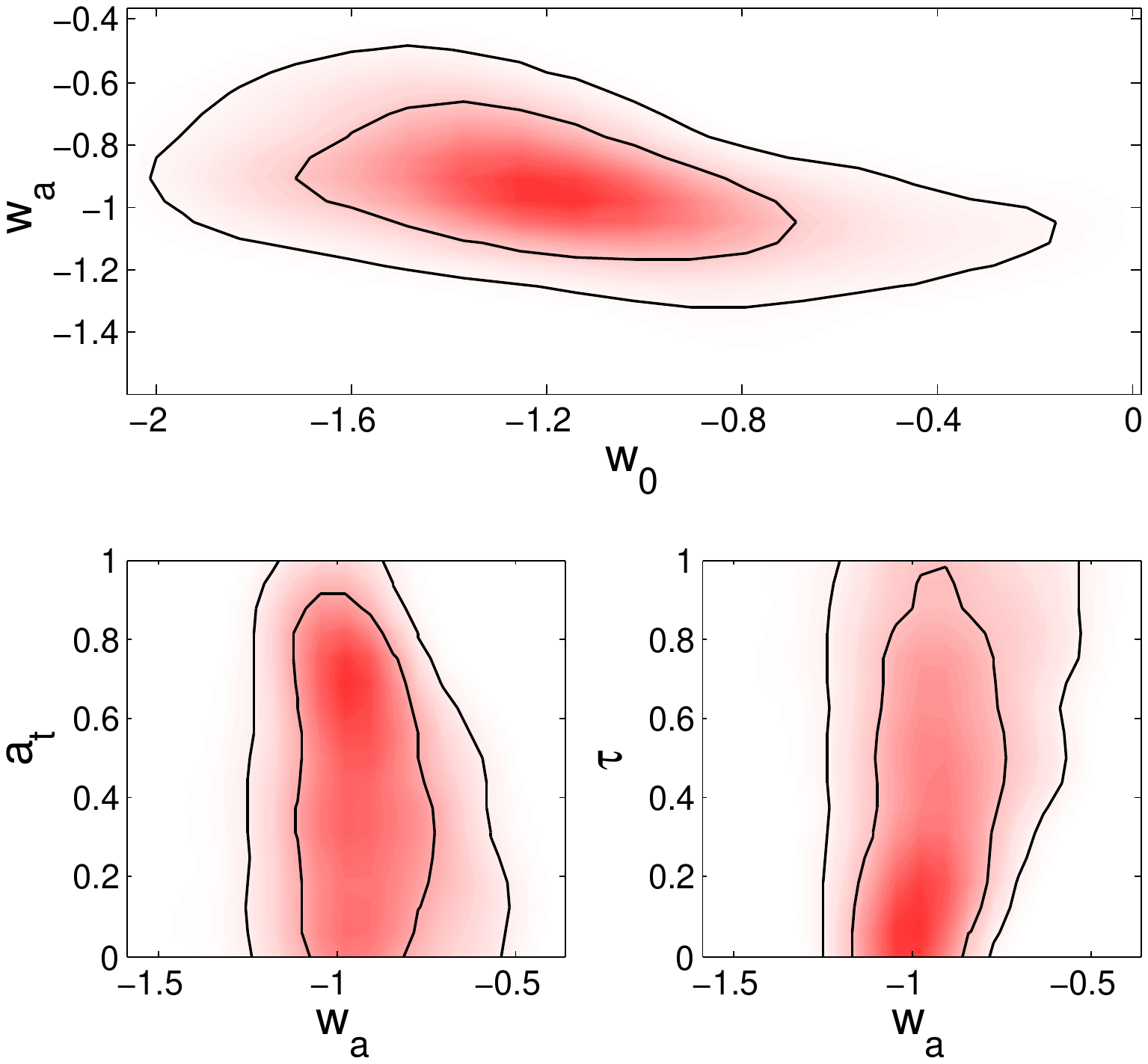}\\
\end{array}$
\end{center}
\caption{Reconstruction of the dark energy equation of state $w(z)$
  assuming the Felice-Nesseris-Tsujikawa paramterisation (left panel), along with their
  corresponding 1D, and 2D constraints with $1\sigma$ and $2\sigma$ confidence
  contours (right panel).  The colour-code indicates the $\ln ({\rm
    Likelihood})$, where lighter regions represents an improved fit;
  the top label in the panel denotes the associated Bayes factor with
  respect to the $\Lambda$CDM model.}
\label{fig:fnt}
\end{figure}

In Figure \ref{fig:fnt} we plot 2D joint constraints, with $1\sigma$ and
$2\sigma$ confidence contours, for the parameters used to describe the
FNT model, and its corresponding reconstruction of $w(z)$.
We observe that the FNT model is in good agreement with a simple
cosmological constant $w(z)=-1$, with current constraints:

\begin{eqnarray*}
({\rm FNT}) &&w_0=-1.19\pm0.32, \quad w_a=-0.94\pm 0.15.
\end{eqnarray*}

\noindent
Given that the best-fit values of $w_0$ and $w_a$ are
very similar, the second term on the left hand side of (\ref{eq:fnt}) 
is almost negligible.
This results in essentially unconstrained values for $a_t$ and $\tau$, and
so $w_a$ becomes the dominant term in the dynamics of $w(z)$.
We have found that the FNT model shares a similar feature common
throughout all the models: $w(z=0)\lesssim w(z\gg 1)$, in agreement with our previous
results.
The best-fit form of $w(z)$ 
presents a maximum value given by $w(a_*)=-0.95$ 
located at $z_*=1/a_*-1=1.59$.
On the other hand, the top label
of Figure \ref{fig:fnt} shows the Bayes factor compared to the $\Lambda$CDM model: 
$\mathcal{B}_{{\rm FNT},\Lambda} =-1.68 \pm 0.35$.
That is, the FNT model improves on the Evidence computed from the
CPL and JBP models, however the inclusion of twice the number of parameters 
makes it significantly disfavored when compared to the cosmological 
constant $w(z)=-1$, and indistinguisable compared to our node-base reconstruction, 
i.e. $\mathcal{B}_{{\rm FNT},z_2} =-0.82 \pm 0.35$.

\section{Discussion and Conclusions}
\label{sec:conclusion}

%
The major task for present and future dark energy surveys is to
determine whether dark energy is evolving in time.  Using the latest
cosmological datasets (SN, CMB and LSS), we have performed a Bayesian
analysis to extract the general form of the dark energy equation-of-state 
parameter, employing an optimal nodal reconstruction where $w(z)$ is
interpolated linearly between a set of nodes with varying 
$w_{z_i}$-values and redshifts.  Our method has the advantage that the number
and location of nodes are directly chosen by the Bayesian evidence. We
have also explored standard parameterisations which include the CPL, JBP 
and FNT models.
We find our results to be generally consistent with the cosmological
constant scenario, however the dark energy does seem to exhibit a
temporal evolution, although very weak.  Besides the cosmological
constant, the preferred $w(z)$ has $w\lesssim -1$ at the present time
and a small bump located at $z\sim 1.3$, whereas at
redshifts $z\gtrsim1.5$ the accuracy of current data is not enough
to place effective constraints on different parameterisations. It is
also interesting to note the presence of a narrow waist in many models, 
situated at $z \sim 0.3$, which is where the constraints on $w(z)$ are tightest.
A dominant feature throughout the reconstruction is the presence of the crossing 
of  the PDL  $w=-1$, obtained within the range $0< z< 0.5$.
Within the GR context, this  phantom crossing cannot be produced by single 
(quintessence or phantom) scalar
fields. Hence, if future  surveys confirm its evidence, multiple fields or additional
interactions should be taken  into account to reproduce this  important  behaviour.  
\\

\begin{figure}
\begin{center}
  \begin{minipage}[c]{0.55\textwidth}
 \includegraphics[trim = 25mm  92mm 18mm 90mm, clip, width=8.5cm, height=5.5cm]{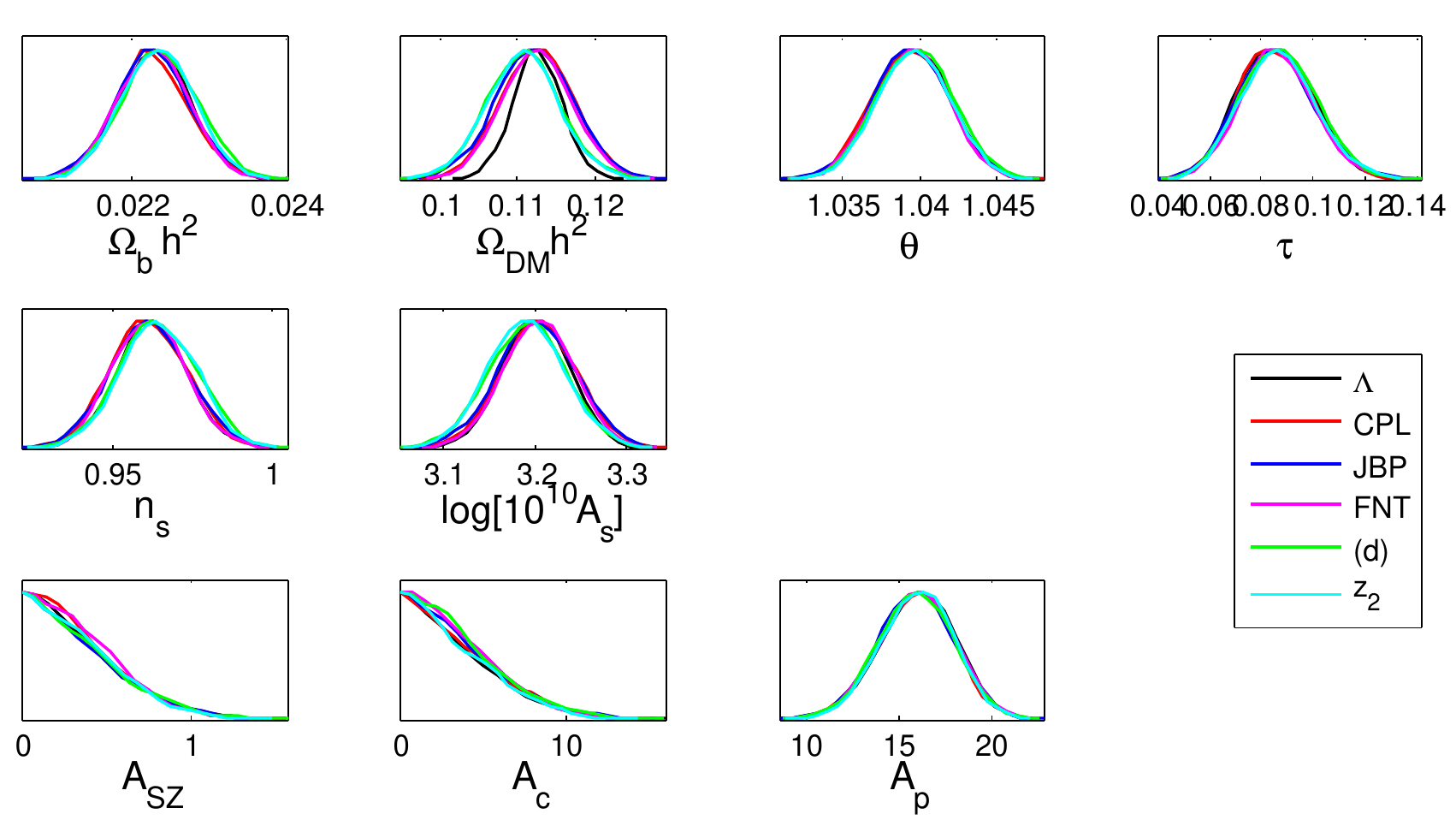}
  \end{minipage} \qquad 
  \begin{minipage}[c]{0.33\textwidth}
  
\begin{tabular}{ccc} 
\cline{1-3}\noalign{\smallskip}
\hline
\vspace{0.2cm}
{\bf Model}&  N$_{\rm  par}$ &  $\mathcal{B}_{i,\Lambda}$   
\\ \hline
\vspace{0.2cm}
$\Lambda$ &			-		&	  	   	$0.0\pm 0.3$    	 	\\
\vspace{0.2cm}	
CPL &				+2		&			$-2.8\pm0.3$				\\
\vspace{0.2cm}
JBP &				+2		&			$-2.8\pm0.3$				\\
\vspace{0.2cm}
FNT &		                +4              &                     $-1.7\pm0.3$                            \\
\vspace{0.2cm}
(d) &					+4		&	 		$-1.6\pm0.3$				\\
\vspace{0.2cm}
 $z_2$&				+6		&			$-0.8\pm 0.3$				\\
\hline
\hline
\end{tabular} 

  \end{minipage}

\end{center}
\caption{Left:
1D marginalised  posterior  distributions of the standard
 cosmological parameters, of each corresponding model  listed  in the right table.                                                                                                                                                                                                                                                                                                                                                                                                                                                                                                                                                                                                                                                                                                                                                                                                                                                                                                                                                                                                                                                                                                                                                                                                                                                                                                                                                                                                                                                                                                                                                                                                                                                                                                                                                                                                                                                                                                                                                                                                                                                                                                                                                                                                                                                                                                 
 Right: comparison of the Bayes factor  $\mathcal{B}_{i,\Lambda}$
 for some selected models with an extra-number of parameters N$_{\rm  par}$.  
Each description  is compared respect to the $\Lambda$CDM model.  }
\label{fig:cosmo}
\end{figure}

All the models considered share a consistent set of primary
cosmological parameters: {$\Omega_{\rm b} h^2$, $\Omega_{\rm DM} h^2$,
  $\theta$, $\tau$, $n_{\rm s}$, $A_{\rm s}$}, in addition to
secondary parameters: $A_{SZ}, A_p, A_c$.  The marginalised posterior
distributions for these parameters are consistent with those obtained
using only the concordance $\Lambda$CDM model.  In Figure
\ref{fig:cosmo}, we plot 1D posterior distributions of the
cosmological parameters for some selected models.  We observe that
their values remain well constrained despite the freedom in
$w(z)$. The only noticeable change is in the dark matter parameter,
where the $\Lambda$CDM model displays the tightest constraints.  
In the same figure we include the corresponding Bayes factors, all of
which are quoted relative to the cosmological constant model.  The
preferred Bayesian description of the $w(z)$ is provided by the
$\Lambda$CDM model, followed by the two-internal-node model $z_2$,
introduced in this work.  It is important to note that the CPL and JBP
models, each with two parameters, are not able to provide an adequate
description for the behaviour of $w(z)$, and are hence strongly
disfavoured using the priors chosen. 
The FNT model with four parameters, from which two of them remained
unconstrained, is significantly disfavoured.
We stress that for the
smallest prior range, the Bayes factor for the JBP model (which does not include the case
$w_0=-1$)  is indistinguishable from that of the $\Lambda$CDM
model, therefore pointing to a possible departure from the
cosmological constant.

\acknowledgments
This  work was carried out largely on the Cambridge High Performance 
Computing cluster,  {\sc DARWIN}. JAV is supported by CONACYT M\'exico.


\end{document}